\documentclass[%
 reprint,
 amsmath,amssymb,
 aps,
 pre,
floatfix,
showkeys
]{revtex4-2}

\usepackage{graphicx}
\usepackage{dcolumn}
\usepackage{bm}

\usepackage{wasysym}
\usepackage{color}
\def\redw#1{{#1}}
\def\greenw#1{{#1}}
\def\magentaw#1{{#1}}
\begin{document}

\title{Properties of pedestrians walking in line without density constraint}

\author{C\'ecile Appert-Rolland}
 \email{cecile.appert-rolland@ijclab.in2p3.fr}
 \affiliation{IJCLab, Univ. Paris-Saclay, CNRS, F-91405 ORSAY, France.}
\author{Anne-H\'el\`ene Olivier}
 \email{anne-helene.olivier@irisa.fr}
\affiliation{M2S, Univ. Rennes, Inria, CNRS, IRISA, Rennes, France}
\author{Julien Pettr\'e}
\email{julien.pettre@inria.fr}
\affiliation{Inria, Univ. Rennes, CNRS, IRISA, M2S, Rennes, France}

\date{\today}

\begin{abstract}

This article deals with the \redw{experimental} study of pedestrian behaviours in some situations of one-dimensional traffic. 
\redw{Participants were pre-organized in a line, and asked } 
to walk either in a straight line with a fast or slow leader, or to form a circle. \redw{The originality of our experimental protocol compared to previous ones on similar situations, is to have left the condition of density free.} While the observed density results from individual decisions in the line case, both density and velocity have to be collectively chosen in the case of circle formation. \redw{Our major findings are the following.} In the case of circle formations, 
we observe that the resulting velocity is very stable among realizations, as if collective decision was playing the role of an average. \redw{For line flows with a slow leader, the same operating point close to the jamming transition}
is chosen as in previous experiments where it was not velocity but density that was imposed. \redw{In the less constrained} line experiments, although participants could choose comfortable headways, they rather stick to short headways requiring a faster adaption - a fact that could come from a “social pressure from behind” \redw{ 
and had previously only been observed at high densities. This could be due to the uncertainty and individual responsibility that stem in less constrained situations.} \redw{ 
Our results open new avenues for future research, as they show that the walking values preferred by humans} in following tasks depend on more factors than previously considered.
\end{abstract}

\maketitle


\section{\label{sec:intro}Introduction}

Empirical studies and experiments on pedestrian dynamics are crucial in order to discover the main principles that guide humans in forming locomotion trajectories while performing interactions. 
They are also important to understand the impact of certain factors on these tasks, such as physiological, psychological or physical factors  \cite{haghani2020empirical,corbetta_t2023}. 
Finally, experiments are an essential step in the development and calibration of simulation models that can reproduce and predict the behaviour of pedestrians - and eventually crowds. This has important applications in architecture, urban design, safety and crowd management. 

In practice, pedestrians move in a two-dimensional world, and they can adjust their trajectory by changing either the direction of their velocity or its modulus (speed), according to a strategy that can be quite complex to determine~\cite{ondrej2010b}.
When carrying out experiments to decipher movement strategies, it can be interesting to limit the observations to a simpler situation, namely a one-dimensional (1D) pedestrian flow in which pedestrians follow each other along a curve without being allowed to pass each other~\cite{subaih_t_c2024}. 
This basic situation makes it possible to understand in particular how people establish relationships between speed and distance within a fixed pair of predecessor/follower.
The situation is easy to reproduce in different experiments, making it possible to compare the behaviour of different groups of pedestrians or to study the effect of external factors.

It is also worth noting that in practice, many 1D flows occur in everyday environments. This may be due to geometric constraints in narrow passageways or when queuing is required. 
Therefore, the study of 1D flows is also important for the design of public facilities where they are likely to occur, since the relationship between speed and distance will determine, for example, the capacity of corridors or exits, as well as evacuation times. 
1D flows can also occur spontaneously, for example between bidirectional flows in corridors, through the formation of lanes, and models based on the 1D interaction between pairs of pedestrians following each other can serve as a basis for more complex situation modelling~\cite{lemercier2012a}.

The literature reports a large number of empirical studies in which participants follow each other without overtaking~\cite{seyfried2005,seyfried2010}.
These papers have opened avenues for understanding the relationships between average density, interpersonal distance (also called headway) and speed~\cite{cordes2023}. The propagation of emergent collective motion patterns, such as stop-and-go waves, has also been studied~\cite{jelic2012a,lemercier2012a}. Further studies have been devoted to understanding the effect of age on the properties of 
1D flow \redw{or 2D flow in quasi 1D geometry such as corridor,} in order to adapt fundamental diagrams (velocity-density relation) to specific populations such as the elderly~\cite{cao2016a,ren2019fundamental} or children~\cite{fang2019experimental}. The influence of environmental factors has also been investigated. Cao \textit{et al.} studied the effect of limited visibility \cite{cao2019dynamic}. Different types of environments were considered that are typically found in transport infrastructure, such as staircases \cite{chen2017pedestrian}, ramps to access vessels \cite{sun2018moving}, or highly constrained environments similar to airplane aisles \cite{huang2018experimental}. More complex paths (e.g. turns) \cite{shi2018state} 
were also considered. Even the sound environment, noise or music, can affect walking \cite{zeng2019experimental}. Yanagisiwa studied the influence of rhythm on the emergent properties of 1D flows \cite{yanagisawa_t_n2012}, which opens up ways to improve evacuation capacity by playing music.

Social parameters~\cite{hu2021social}
as well as cultural parameters \cite{chattaraj_s_c2009a,gulhare2018comparison} were shown to have a measurable but small effect on \redw{(quasi-)}1D fundamental diagrams. Generally, experiments are specific to one country and its 
population~\cite{zhang_h2024}. 
Locomotion plays a much more dominant role. 1D flows are particularly well suited to study the relationships between step length and frequency at different distances, speeds or in different groups of people \cite{jelic2012b,zeng2018experimental,cao2018stepping,ma2018pedestrian,ren2021flows}. It has been shown that the stepping laws observed in line flows are significantly different from those for isolated individuals, and that collective dynamics can induce step coordination phenomena \cite{jelic2012b,seyfried2005}. 

A common feature of most of these previous works has been to constrain the density of pedestrians, or to control their headway, in order to observe its effect on other characteristics of the flow, whether global or local: global flow, individual speed, stepping activity, and so on. For example, in \cite{jelic2012a} it was possible to have controlled densities ranging from 0.31 to 1.86 ped/m \redw{ by prescribing the circular trajectory on which a fixed number of pedestrians had to walk}.

In this paper we report on 1D flow experiments designed in such a way that the density is not prescribed a priori. Instead, the density results from the spontaneous behaviour of the participants.
The situations we considered to achieve this lack of density constraint were as follows.
In a first set of experiments, participants had to walk in a straight line and follow each other. Thus, their speed is constrained by their leader, but they are free to choose their distance from their predecessor, as it would be in a queuing system~\cite{yanagisawa2013walking}. 
In a second experiment, we asked participants to form a circle with no indication of its radius.
In both experiments, one might wonder what density is spontaneously chosen from the participant dynamics, and whether or not the resulting stationary flow follows the same fundamental diagram as in the case of density-constrained flows.

\section{\label{sect_exp}The PEDINTERACT experiments}

\subsection{Experimental protocols}

\begin{figure*}[ht] 
   \centering
   \includegraphics[width=0.60\columnwidth]{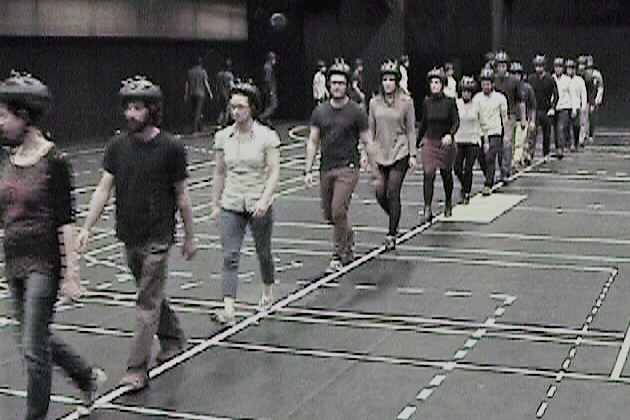}
   \includegraphics[width=0.60\columnwidth]{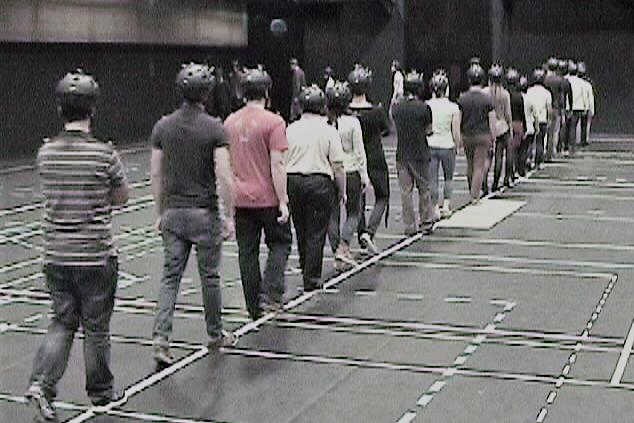}\\
   \includegraphics[width=0.49\columnwidth]{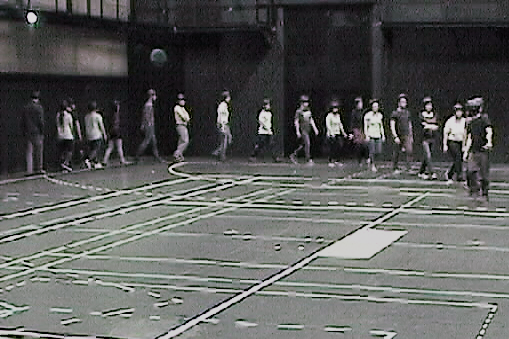}
   \includegraphics[width=0.49\columnwidth]{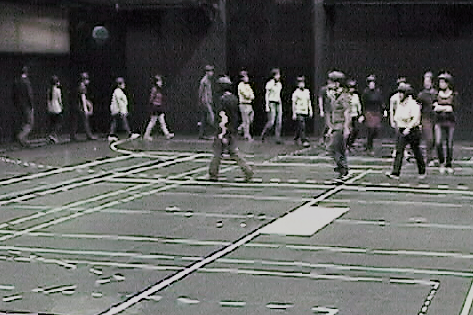}
   \includegraphics[width=0.49\columnwidth]{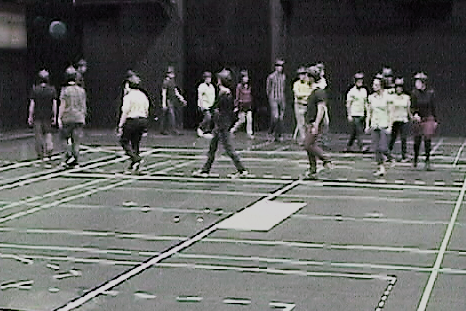}
   \includegraphics[width=0.49\columnwidth]{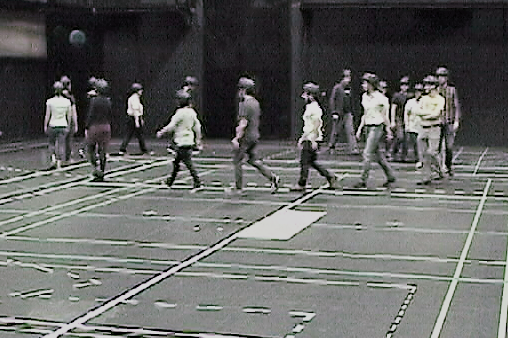}
   \caption{\textbf{Top} Pictures taken during the line experiments. Left / Right: the first participant walks along a line at normal / slow speed. 
   \textbf{Bottom}: Formation of a circle by a group of participants.}
   \label{fig_expe_types}
\end{figure*}

The general aim of the experiments carried out in the PEDINTERACT project 
was to study the spontaneous choices that pedestrians make in their walking when \redw{following in groups and when} density or speed constraints are relaxed. \redw{We recruited 74 participants total, 48\male~and 26\female, aged $31.8 \pm 10.8y$.} More specifically, two types of experiments were carried out:
\begin{itemize}
    \item \textbf{Line experiments:} participants were asked to cross a room following a line drawn on the floor (see Fig. \ref{fig_expe_types}, top row). They were not allowed to pass each other. We studied two conditions, one in which the first participants had no further instructions on how to cross the room, called \textit{line-free}, and one in which the first participant was asked to walk slowly, thus limiting the speed of all participants behind him, called \textit{line-slow}.  
    \item \textbf{Circle experiments:} \redw{participants were prepositionned in a line against the wall of our experimental room,} and were asked to follow each other to form a closed circle (see Fig. \ref{fig_expe_types}, bottom row). They were not allowed to pass each other. Neither the radius nor the center was given. 
\end{itemize}

What these two types of experiment have in common is the absence of control over the density of the participants, who are free to adjust locally the distance separating them from their predecessor, i.e. their headway. Density, whether global or local, is thus the sum of each participant's contribution. Speed was also left uncontrolled, except in the \textit{line-slow} experiment, where we forced participants to walk at a slow speed by asking the first participant to walk slowly. In all other cases, speed was a free parameter. Of course, since participants could not walk faster than their predecessor because they were asked not to overtake, the predecessor's speed can be considered as an upper bound.  

\subsection{Experimental setting}

\begin{table}
\begin{center}
{\small
\begin{tabular}{|l|l|l|}
\hline
Experiment type & Number of trials & N. of participants \\
\hline
\textit{Circle} & 6 & 18 \\
\textit{Line-free} & 7 & 31 to 36 \\
\textit{Line-slow} & 3 & 33 to 36 \\
\hline
\end{tabular}
}
\caption{Number of trials and of involved participants for each type of experiment}
\label{tab1}
\end{center}

\end{table}

The experiments~\cite{appert-rolland2020b} were carried out in a large sports hall which is part of the Immermove platform of the M2S laboratory (\redw{University Rennes 2}, France). Two sets of experiments were performed on $30^{th}$ and $31^{th}$ of March 2016, with 36 adults participating each day. Our study complied with the Declaration of Helsinki and was approved by our local ethics committee.

Each experimental protocol was carried out on both days - with different participants - and repeated several times to obtain some statistics. We call each repetition of an experiment a \textit{trial}. Groups of up to 36 participants were involved in each trial of the \textit{line-free} and \textit{line-slow} experiments. Half groups of 18 participants were involved in each trial of the \textit{circle} experiments. 
Table \ref{tab1} shows the number of trials we ran for each type of experiment. Note that the order of participants in each trial was randomised, as was the composition of the sub-groups of 18 people for the circle experiment. 

We collected the movement data using a Vicon system consisting of 26 cameras that tracked 4 reflective markers glued to the top of the head on a helmet worn by the participants.
Markers were also placed on the shoulders of the participants.

\subsubsection{Line experiments}

\begin{figure}[ht] 
   \centering
   \includegraphics[width=0.49\columnwidth]{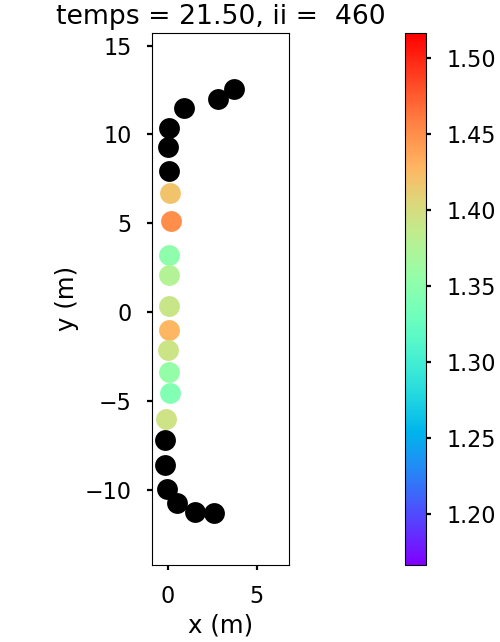}
   \includegraphics[width=0.49\columnwidth]{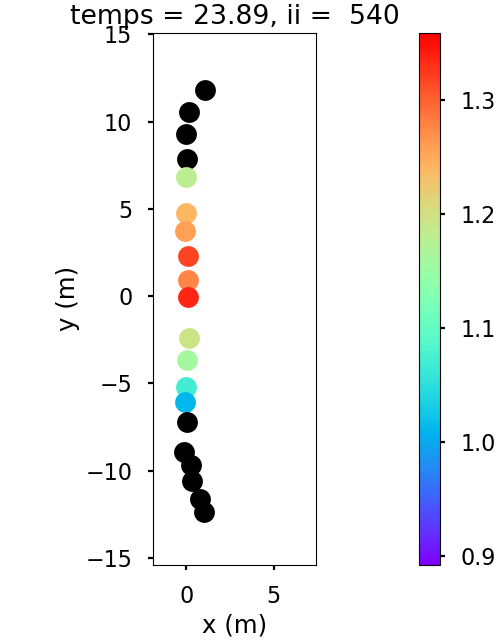}
   \caption{Instantaneous positions of participants during the line-free experiment. Participants are represented by disks colored after their velocity (colorbar, in m/s). Distances are expressed in meters.
   }
   \label{fig_snapshotl}
\end{figure}

Fig.~\ref{fig_expe_types}-top shows pictures of the line experiment, Fig.~\ref{fig_snapshotl} gives a top view of the instantaneous positions of participants at given times. The line was approximately $20m$ long, and one can see participants approaching the beginning of the line and leaving it, forming curved trajectories at extremities \redw{(seen in Fig.~\ref{fig_snapshotl}).} \redw{The way participants joined and left the central line prevented them to perturb the motion of participants in the line}. Though, in order to avoid boundary effects, we have retained for our analysis only the central part of the line, far from extremities (approximately the 14 central meters of the line). In Fig.~\ref{fig_snapshotl} participants in (resp. outside) the measurement area are represented by colored (resp. black) disks.
Plots of the whole set of trajectories for given realizations of the line-free and line-slow experiments can be found in Fig.~2 of Section~II in Suppl. Mat.

While the number of participants was almost the same in all line experiments,
the duration of the \textit{line-free} experiments (from 45 to 60 seconds) was significantly shorter than that of the \textit{line-slow}
experiments (from 77 to 88 seconds), as expected.

\subsubsection{Circle experiments}

Fig.~\ref{fig_expe_types}-bottom shows pictures of the early stages of a circle experiment, with the gradual formation of a circle. There was no indication given to the participants about the size of the circle to be formed.

Once the circle is formed, Fig.~\ref{fig_snapshotl} shows the participants' trajectories for one given realization (plots for other realizations can be found in \redw{Fig.~3 of} Section~II of Suppl. Mat.), as well as the instantaneous positions of the participants at a given time with speed being colour coded.

It can be seen that the participants were able to complete the task correctly, forming a circle quite accurately - an observation that will be quantified later.

\begin{figure}[ht] 
  \centering
   \includegraphics[width=0.7\columnwidth]{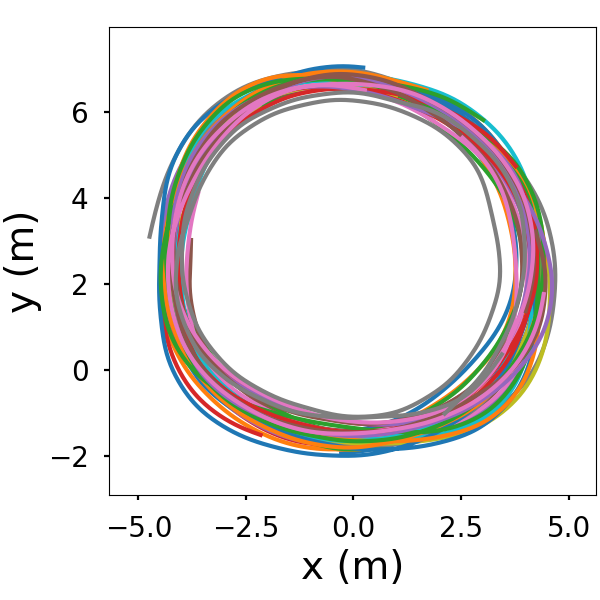}
   
   \includegraphics[width=0.7\columnwidth]{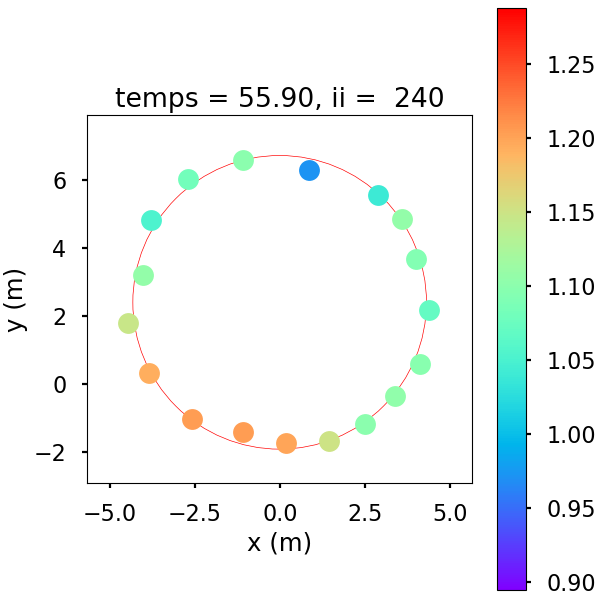}
   \caption{For one realization of the circle experiment (duration 50s):
   (Top) Trajectories of all participants; 
   (Bottom) Snapshot at a given time of the experiment.
   Participants have been replaced by a disk
   colored after their velocity (colorbar, in m/s). Distances are expressed in meters.
   The fitting circle is also shown.}
   \label{fig_snapshotc}
\end{figure}

\redw{For measurements which had to be performed in the stationary state, such as the fundamental diagrams of section~\ref{sect_fd}, the most restrictive time interval obtained from 3 independent criteria was used: The circle had to be formed, meaning that all pedestrians were inside the black circle
shown in Fig.~\ref{fig_circle_formation}. Besides, initial and final transients both in velocity and headway profiles were determined by visual inspection and suppressed.}

\redw{
The rotation direction, though not imposed, was probably induced by the initial condition: pedestrians were waiting in a line along the walls,
and the location of the person giving the instructions may have oriented them.
Among 6 circle experiment, 5 gave a circle turning clockwise, an observation which is probably not significant. 
To make the data analysis more easy, all data were symmetrized to become anti-clockwise.}

\subsection{Postprocessing}

The VICON software was used to reconstruct the 3D trajectory of each marker,
and associate the markers with a given participant~\cite{lemercier2011b}.

\subsubsection{Head and shoulder positions}

At each time frame, we reconstruct two independent positions for each participant.
One is the average position of the head markers, the other is the average position of the shoulder markers.

Markers are sometimes occluded, especially those on the shoulders.
When a head marker is occluded, the VICON software allows its position to be reconstructed from the position of the other head markers, provided that enough of them are visible.
However, this is not always possible.
We only keep the data if all the marker positions of one type (head or shoulder) are known (directly or from reconstruction).

Shoulder-based positions are more likely to be missing due to occlusion,
but is a priori more accurate than the head position, as the head can move independently of  the whole body, at least to some extent.
A secondary aim of this paper is to estimate the discrepancy between the two types of measurements.
Throughout the paper we will therefore investigate how much our results can vary depending on which part of the body is tracked.

\subsubsection{Step filtering}

Participants' steps and sway induce some oscillations around the mean trajectory, which should be filtered out before, for example, estimating the distance between two participants.
To make the data easier to handle, we have kept only 25 frames per second from the original 120 frames per second.
We want to filter out steps with a period of about $2$s \redw{(for two steps, or equivalently one complete step cycle), i.e. with a frequency of $0.5$ Hz.}
\redw{To do this we used a 2nd order low pass Butterworth filter with a sampling frequency of $25$ Hz and a cut-off frequency of $0.42$ Hz, slightly below the step frequency we want to attenuate, as justified by Fig.~1 in the Suppl. Mat.
The advantage of the Butterworth filter is that the frequency response is as flat as possible in the low-pass band.}

As explained above, trajectories can be interrupted from time to time due to occlusion.
Applying the filter directly to short trajectories can cause strong boundary effects. Also, it is useless to filter out steps if the trajectory piece is too short compared to the step period.
To avoid this, we initially only kept trajectory segments longer than 4 seconds.
Also, before applying the low-pass filter, we applied some reflections to the trajectory pieces as explained in Section~I of Suppl. Mat.

\section{Spontaneous choice of headway and velocity in density-free 1D flows}
\label{sect_fd}

At any time, it is possible to define the instantaneous speed of a given participant as the derivative of his filtered position.
It is interesting to relate the norm of the instantaneous velocity (speed) to the instantaneous distance of the participant from his predecessor (headway) or to the instantaneous individual density defined as the inverse of the headway~\footnote{Some authors define the density from a headway which is the average of the distance with the predecessor and the follower~\cite{seyfried2010}. Here, assuming that it is the distance {\em ahead} of the participant that predominantly determines his speed, we have chosen to consider only the headway ahead.}.

\redw{
Fundamental diagrams under the form of 2d histograms can be obtained for each realization of the experiment (see. Section III in Suppl. Mat. for an example for a particular realization of the circle experiment and of the line experiment).}

\begin{figure}[ht] 
   \centering
   \includegraphics[width=1\columnwidth]{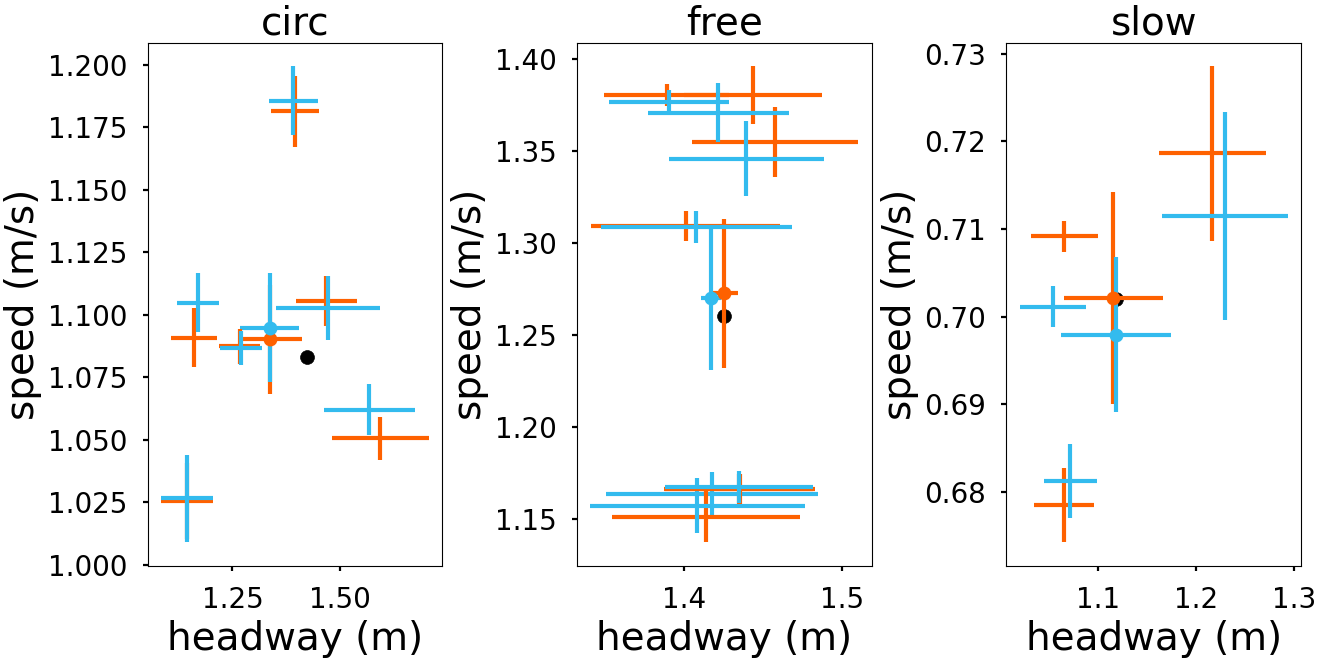}
   \caption{
   Fundamental diagram for each experiment type (circ on the left, line-free in the middle, line-slow on the right) and comparison of 
   \redw{head (orange) and shoulder (cyan) tracking.}
   The instantaneous individual velocities and headways have been averaged for each realization, with the SEM represented by error bars.
   The average over all realizations is indicated by a disk (orange or cyan depending on the head/shoulder tracking; black when weighted by the duration of the experiment, for head-tracking only).
   }
   \label{fig_fd}
\end{figure}

An average value and its associated SEM - Std Error of the Mean - is estimated from the (speed, headway) values of each experimental realization.
All these values are reported on the Figures~\ref{fig_fd} for the different types of experiments. A second average is then taken over the realizations.

These measurements can be used as a benchmark to evaluate the differences between head and shoulder tracking.
In fact, both types of tracking give quite similar results, although the discrepancy is greater for the slow line. Indeed, it is expected that swaying is more important when participants are forced to walk slowly, which could explain the difference.

\begin{figure}[ht]
   \centering
   \includegraphics[width=1\columnwidth]{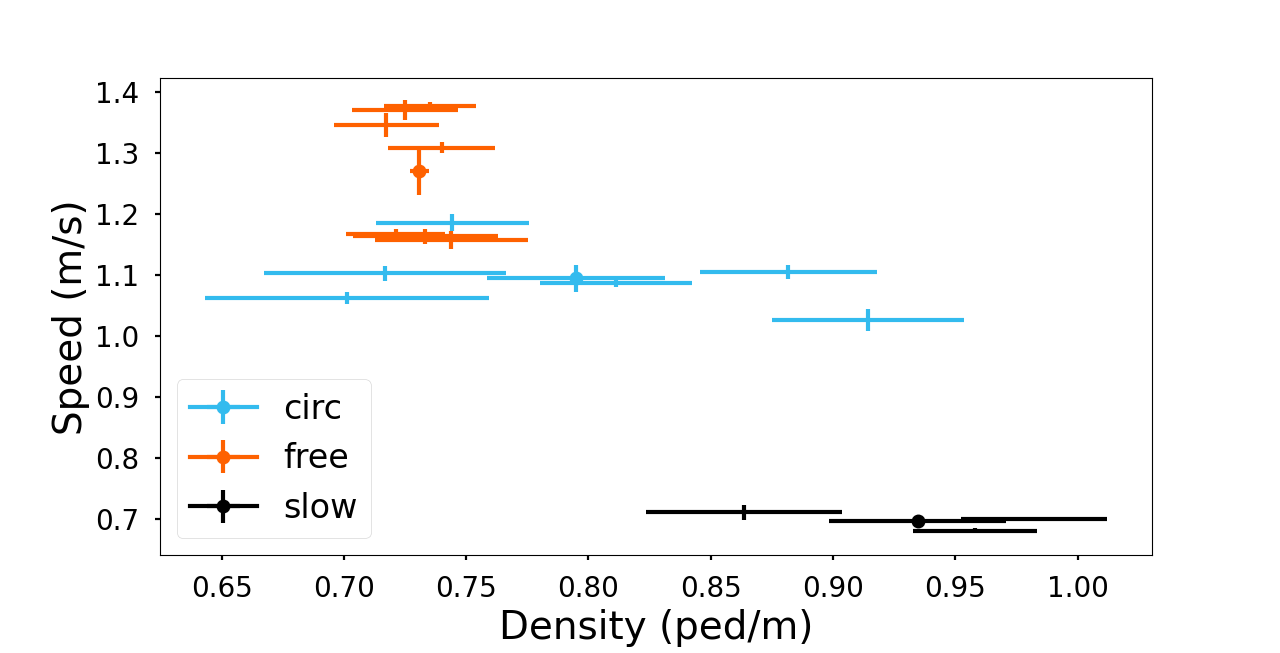}
   
   \includegraphics[width=1\columnwidth]{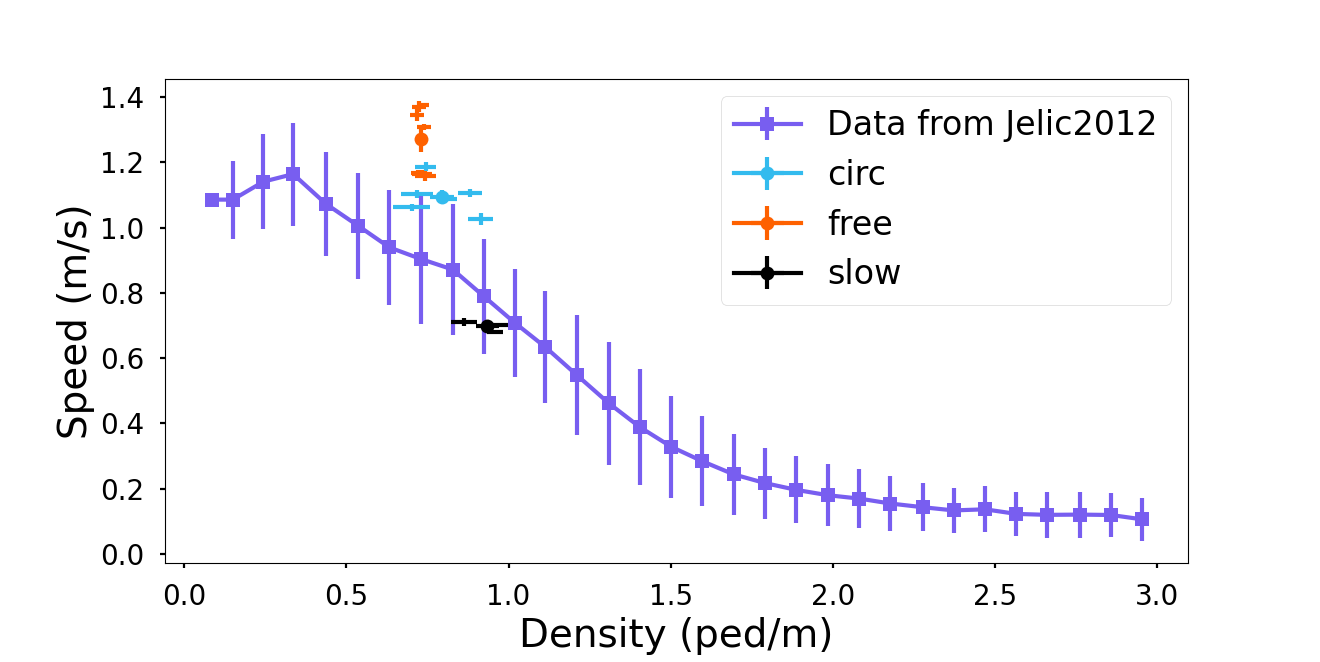}
   \caption{Velocity as a function of density for all experimental realizations, \redw{based on shoulder tracking, for the circle (cyan), line-free (orange), and line-slow (black) experiments.}
   Error bars are here error on the mean.
   At the bottom, the same data are superimposed with the data \redw{by Jelic et al from } Fig.~3
   in~\cite{jelic2012a}\redw{, in which the global density was imposed}.
   For the Jelic's data, bars give the standard deviation.
   }
   \label{fig_vdall}
\end{figure}

In order to make it easier to compare the results obtained for the different types of experiment, 
the same data as in Fig.~\ref{fig_fd} are collected in Fig.~\ref{fig_vdall}.

Let us first compare the results of the circular (red-magenta) and line-free (blue-black) experiments (Fig.~\ref{fig_vdall}-top).
It is striking that the density is almost the same for all the line-free realizations (with varying speed), whereas for the circle experiment it is the opposite: the speed is almost the same for all realizations and the density varies.

Several hypotheses can be put forward to explain this difference.
In both types of experiment, the participants were free to choose their heading (or local density) and their speed. In the circle, however, the periodic boundary conditions mean that all participants have the same average speed. The resulting speed is therefore the result of a tacit negotiation, which, like averaging, can lead to more stable values.

It could be argued that in the line-free experiment, people were also free to choose their speed, provided it was lower than that of their predecessor.
However, in this case there is no possibility of visual negotiation, and one may wonder if the higher speeds observed are not due to a social pressure resulting from the fact that each participant knows that his own speed will limit the speed of all his followers.
\redw{Actually, a more careful observation of the spatio-temporal plots (See Fig.~\ref{fig_stlf} in this paper or Fig.~7 in Suppl. Mat.) or of the individual trajectories (Fig.~11 in Suppl. Mat.) shows
that in all realizations of the free line experiment, the average velocity of pedestrians tends to decrease with time. Still, if a lower velocity was more comfortable, pedestrians could have walked more slowly from the beginning, as their velocity is only upper bounded by their predecessor. This could indicate some kind of social pressure that limits abrupt velocity changes. In any case, we see in Fig.~\ref{fig_vdall}-top that the velocity is strongly varying 
\redw{between trials}, while the mean headway is much more constant over time.
Note however that this almost constant mean value does not imply that all participants walk with the same headway, but rather that the average for each realization scans the whole distribution of possible headways, as illustrated by Fig.~12 in Suppl. Mat.}

The fact that we observe lower speeds in the circle case \redw{(compared to line-free experiments)} could be related to the fact that curved trajectories are more difficult to follow from a locomotion point of view than straight lines.
However, lower speeds could also be due to 
the need to make a collective decision about the circle, hence some attentional demand
which can slow down the whole movement.

\redw{
In Fig.~\ref{fig_vdall}-bottom we have superimposed our data to those from a one-dimensional flow experiment carried out in 2010 by Jelic et al~\cite{jelic2012a} in which the overall density was imposed: a (controlled) number of participants followed each other along a predefined circle. This figure therefore allows us to compare the characteristics of the movement of a queue when the density is controlled as in~\cite{jelic2012a} or left free as in our case.
}

It should be noted that \redw{the speeds observed in the circle experiments
of our paper, though lower than the ones of line-free experiments, still are higher} than those of~\cite{jelic2012a}
where participants had \redw{also to walk along a circle, but close to} a wall.

Regarding the variability of density in the circle experiment, it appears that participants choose it very early in the formation of the circle, a fact that will be discussed further in section~\ref{sect_circ}.
This may lead to more atypical density choices.

In the line-slow case, the speed is strongly constrained by the low speed of the leader.
As a result, the speed is almost constant in all realizations.
Note that the participants could have chosen to maintain some large and comfortable headways.
However, this is not the case.
If we compare with the fundamental diagram obtained in~\cite{jelic2012a}
along a fixed circular path, we find that the line-slow regime coincides with these older data.
This is quite surprising as the decision process is quite different.
In~\cite{jelic2012a}, the global density (and thus the average walking speed) is fixed, and participants can only collectively adjust their speed.
Here it is the speed that is fixed (at least strongly upper bounded), 
and participants can choose their headways individually.
The fact that the same result is obtained probably indicates that some
automatic behaviour associated with walking in constrained flows.
It may also be that in the line-slow experiment, although participants could choose 
arbitrarily large distances, there is again a social pressure from behind that prevents them from doing so.

\begin{figure}[ht] 
   \centering
\includegraphics[width=1\columnwidth]{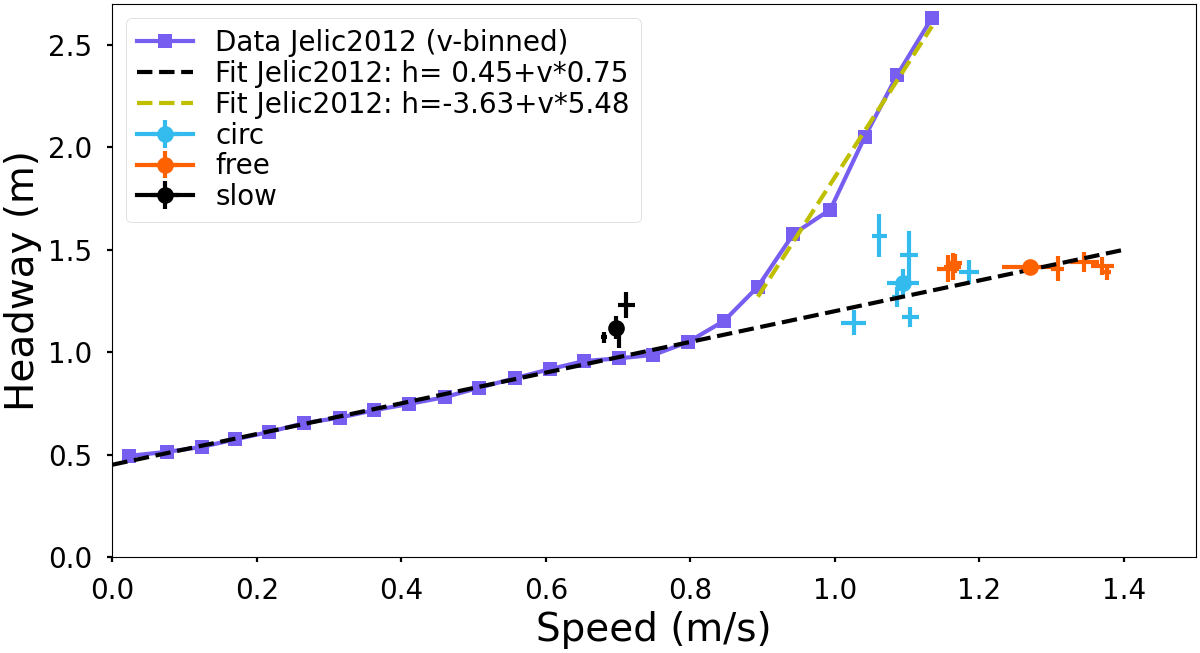}
   \caption{Velocity as a function of the headway for all experimental realizations, for the 
   \redw{circle (cyan), line-free (orange), and line-slow (black) experiments. 
   \redw{For comparison, the} data of Fig.7
   in~\cite{jelic2012a}, in which the global density was \magentaw{imposed, are displayed with blue squares.
   The dashed black line (resp. yellow) is a fit of the low (resp. intermediate) speed data of~\cite{jelic2012a}.}
   }}
   \label{fig_vh}
\end{figure}

In \cite{jelic2012a} we \redw{had} found that the jamming transition corresponds to a change in the typical time scales involved (slope of the curve~\magentaw{ with blue squares} in Fig.~\ref{fig_vh}). This allowed us to distinguish a weakly constrained regime at intermediate densities and a strongly constrained regime at high densities (small headways).
\magentaw{While the weakly constrained regime is associated to a time scale around $5$s 
typical of reaction times when we are not prepared to an event, the strongly constraint regime is associated to a much smaller time scale of $0.75$s, as shown by the fit of low speed data of~\cite{jelic2012a} in Fig.~\ref{fig_vh}. The linear relation $h=0.45+v*0.75$ can be interpreted~\cite{seyfried_p_s2010} as the sum of a minimal distance $h_0=0.45m$ that pedestrians want to keep even at zero speed, and of a velocity dependent contribution to the safety distance. The minimal distance $h_0$ includes the physical exclusion length due to body size and some minimal interdistance that people would respect anyway if not pushed, and that may depend on the cultural background~\cite{chattaraj_s_c2009a}.
Further distance is taken from the predecessor when velocity increases. This second term can be interpreted as the distance traveled during a reaction time of $0.75$s if the pedestrian would keep his/her current velocity.}

\magentaw{It appears that the data of our new experiments, }
 although the constraints are relaxed in many ways \magentaw{ and quite large headways are involved}, 
still \magentaw{mostly align with} the high density regime of~\cite{jelic2012a}. 
This would mean that in the new experiments of the present paper, participants remain vigilant and stay on a fast 
\magentaw{reaction } branch, even though they would be free to choose a more relaxed pace and larger distances. One might wonder if this could be a sign of social pressure: participants have more freedom, but their decisions have collective effects, hence a higher level of attention.

\section{Wave propagation in line experiments}
\label{sect_line}

\begin{figure}[ht] 
   \centering
   \includegraphics[width=0.8\columnwidth]{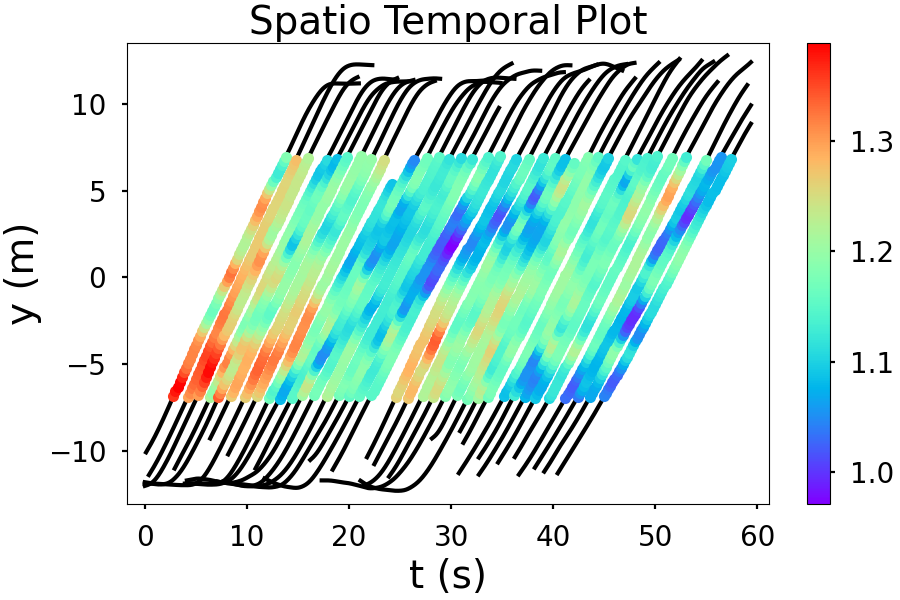}
    \includegraphics[width=0.8\columnwidth]{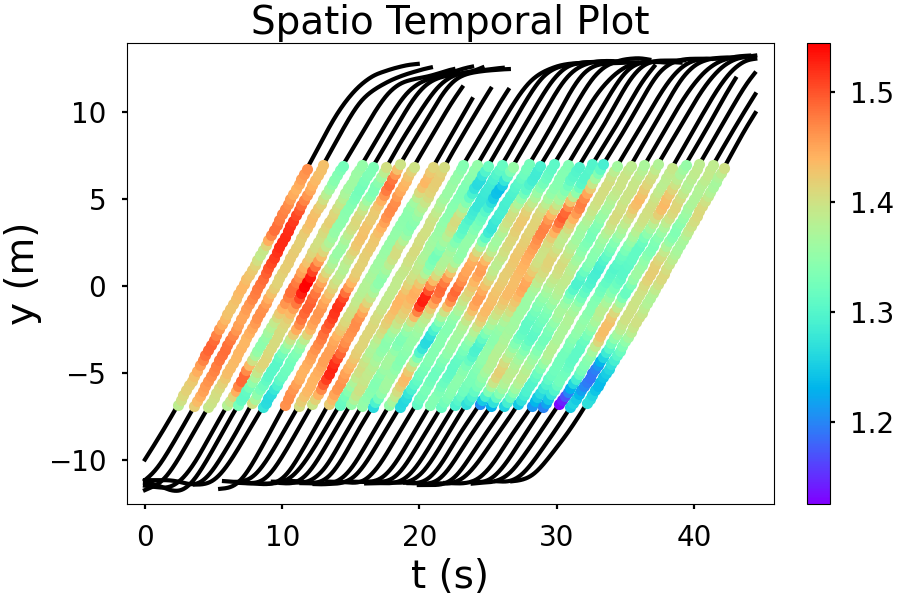}
   \caption{Spatiotemporal plot, for two realizations of the free line experiment.
   In the measurement area, individual trajectories are colored according to the velocity of the participants (scale in $m/s$ given by the color bar).
   }
   \label{fig_stlf}
\end{figure}

\begin{figure}[!h] 
   \centering
    \includegraphics[width=0.8\columnwidth]{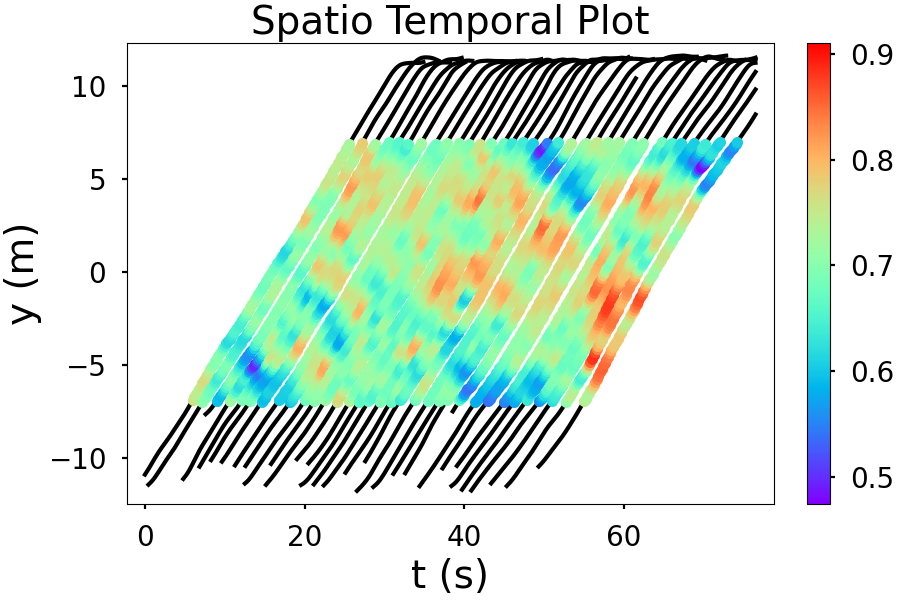}
    \includegraphics[width=0.8\columnwidth]{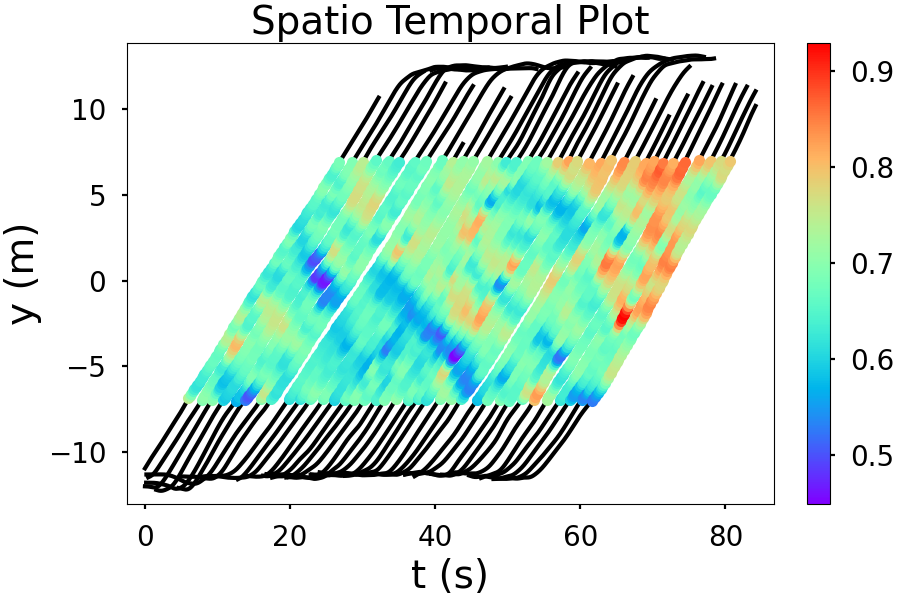}
   \caption{Spatiotemporal plot, for two realizations of the slow line experiment.
   In the measurement area, individual trajectories are colored according to the velocity of the participants (scale in $m/s$ given by the color bar).
   }
   \label{fig_stls}
\end{figure}

Spatio-temporal plots are useful tools to reveal global structures in the flow.
Figures \ref{fig_stlf} and \ref{fig_stls} show examples of such spatio-temporal plots for the free and slow line experiments. Individual trajectories are coloured according to the speed of the participants, revealing the propagation of velocity waves. All other spatio-temporal plots can be found in Section~IV of the Suppl. Mat.

\begin{figure}[ht]
  \centering
  \includegraphics[width=0.90\columnwidth]{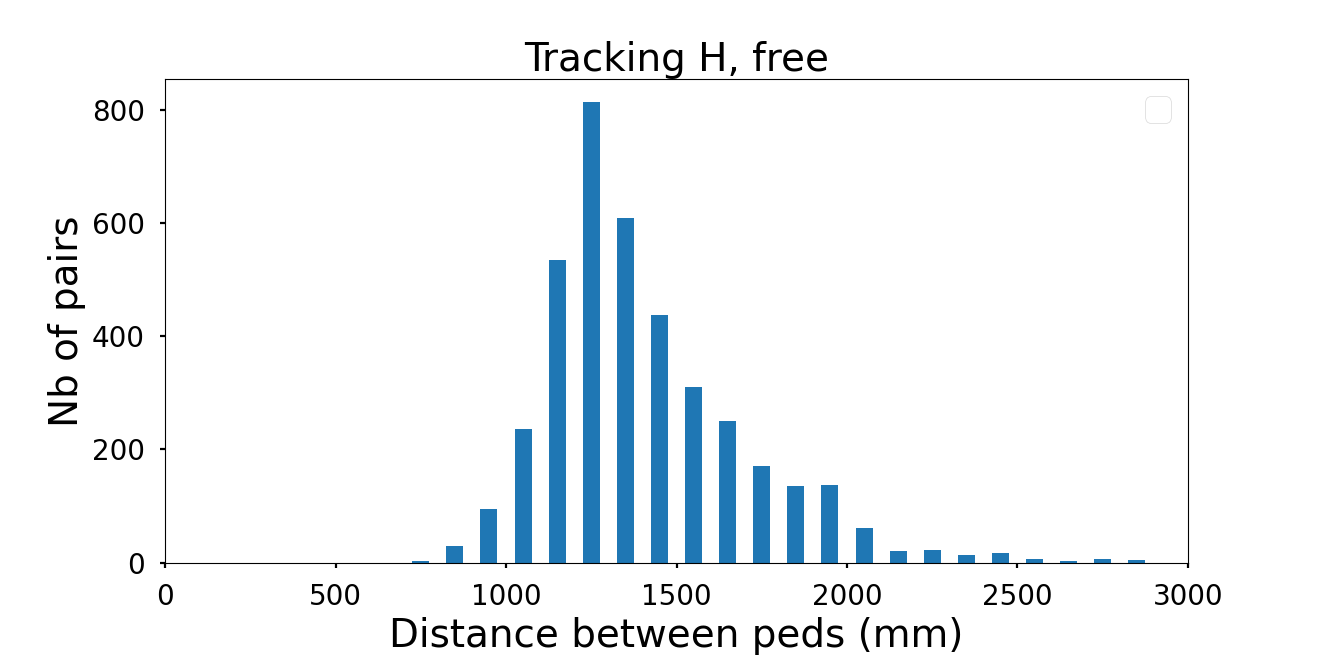}

  \caption{\redw{Number of pairs of successive pedestrians, as a function of the distance $r$ separating the two participants of the pair. The histogram is computed for all times and all pairs of participants, and averaged on all realizations of the free line experiment, with tracking based on heads.}
  }
  \label{fig_nbin_v}
\end{figure}

\begin{figure}[!ht] 
  \centering
  \includegraphics[width=0.90\columnwidth]{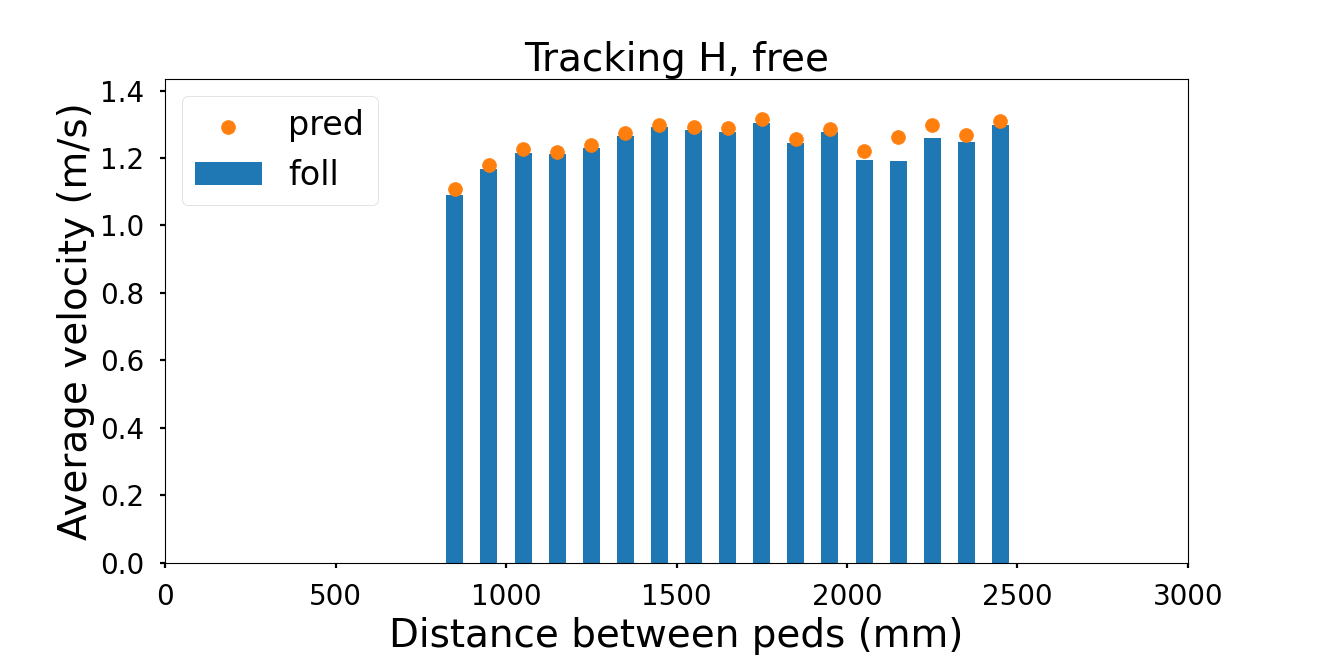}
   \includegraphics[width=0.90\columnwidth]{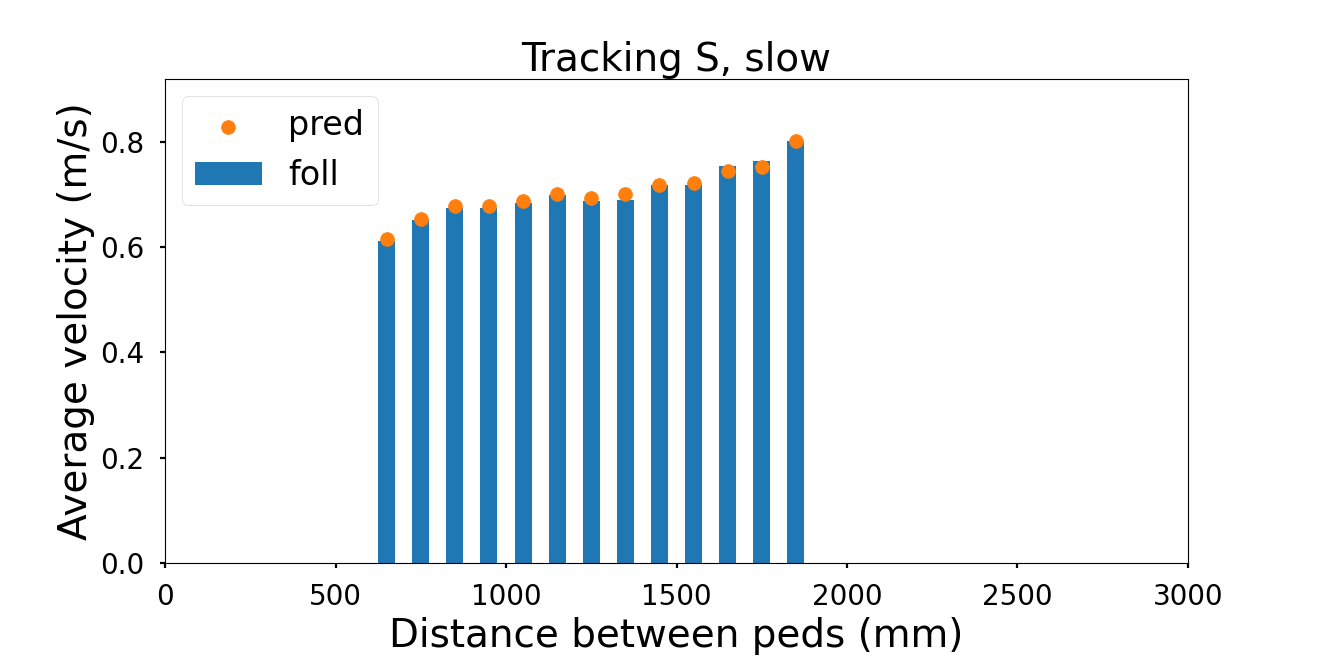}

  \caption{\redw{Velocity of the predecessor and follower in a pair of successive pedestrians, computed for all times and all pairs of participants, in the case of the free line (left) or slow line (right). The distance $r$ is the distance between the participants of each pair. Tracking is based on heads (left) and shoulders (right) in order to avoid too much swaying in the case of the slow line.}
  }
  \label{fig_vmoy_pair}
\end{figure}

Comparing the figures for both types of line experiments shows that the direction of wave propagation is not the same in the free and slow cases.
In the slow line case (Fig. \ref{fig_stls}), the waves propagate backwards, as is usually the case for the stop-and-go waves observed at higher densities~\cite{lemercier2012a,fehrenbach2015}. The same phenomenon of backward propagating stop-and-go waves is also observed for car traffic~\cite{nakayama2016}. The important ingredient seems to be that participants are prevented from walking as fast as they would like by the presence of a slow-moving predecessor.

In contrast, in the free-line case (Fig. \ref{fig_stlf}), although the signal is more noisy, the waves seem to propagate in the same direction as the participants. However, the speed of the waves is different from that of the participants.

Again, this is consistent with previous experiments performed at free flow densities in a circular corridor, where density/velocity anomalies were more or less advected with the flow, albeit at a slightly lower speed~\cite{motsch2018}.

\redw{As waves stem from interactions between successive pedestrians, we studied the velocity correlations between predecessors and followers.}
\redw{More precisely, we consider pairs of successive pedestrians separated by a distance $r$.
The number of such pairs is given in Fig.~\ref{fig_nbin_v} for the free line with head tracking - other cases are given in Fig~9 of Suppl. Mat.}

\redw{
First we observe that participants tend to slow down when they are close to others (cf Fig.~\ref{fig_vmoy_pair},  and Fig.~10 of Suppl. Mat.) and
the velocities of the predecessor and follower are the same on average.
Surprisingly it can be noted that for the same headway, velocities are quite different in the slow and free line.
A possible explanation would be that pedestrians actually look after more than one predecessor and can evaluate the global dynamical state of the line - a hypothesis that could be tested through further experiments. 
}

\redw{
Besides in the slow case, an almost linear relation can be observed between distance and speed. 
One possible explanation is that when their headway are lower (larger) than in the local average value, pedestrians slow down (accelerate) in order to relax towards a mean behavior.
Indeed, by contrast with the fundamental diagrams of section~\ref{sect_fd}, here we scan fluctuations around the mean.  
}

\greenw{Some general trends can be observed. First (see Fig.~13 from Suppl. Mat.), velocities tend to slightly increase when pedestrians get closer to the end of the line - probably due to a phenomenon that we call in France "the horse feeling the stable".
}

\begin{figure}[ht] 
   \centering

\includegraphics[width=0.45\textwidth]{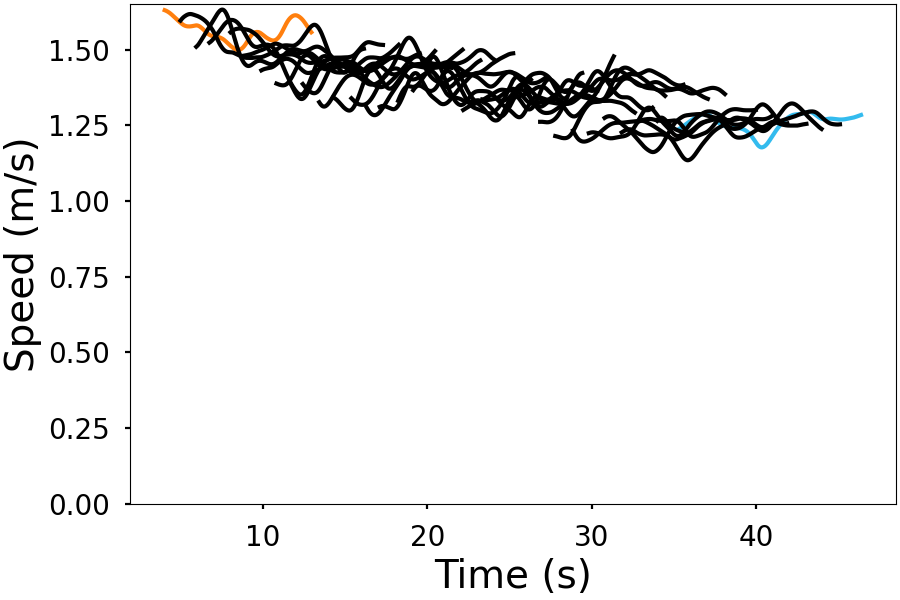}
   \caption{Speed as a function of time for all trajectories (based on head tracking), in one given realization of the free line experiment.
The trajectory of the first (resp. last) pedestrian is in orange (resp. blue).}
   \label{fig_v_t}
\end{figure}

\greenw{
As we already mentioned, a stronger effect is the decrease of velocity as a function of time, on time scales larger than the one required for an individual to cross the hall (see Fig.~\ref{fig_v_t} and Fig.~11 from Suppl. Mat.). This could be due to some weariness, but more probably to the fact that followers can only choose to go more slowly than their predecessor, but not faster, leading to a global decrease of velocity on long time scales. In particular, in Fig.~\ref{fig_v_t}, the entrance of a few slower pedestrians around time $t=30$s leads to a sudden decrease of the average velocity that will persist until the end of the experiment.}

\section{Analysis of the circle experiments}
\label{sect_circ}

Participants were asked to walk in a circle, but it was not obvious a priori how accurately they would perform this task. Figure \ref{fig_snapshotc} shows that the participants successfully completed the task. We can see that the shape of their trajectory is close to a circle with a diameter of about $8m$. Also, the characteristics of the formed circle seem to be quite constant over time. 

We further quantified our visual impressions by considering the 6 realizations of the circle experiment and fitting a circle to the participants' positions at each time step. We looked at the displacement of the center of the fitted circle, the variations in its radius, and the distance of each participant from the mean radius, as we will now explain.

For each experiment, we removed the transients at the beginning and at the end
of each experiment and performed our analysis in the steady state. The steady-state duration for each of the 6 realizations of the experiments was:
71, 50, 25, 24, 24 and 10 seconds.

\begin{figure}[ht] 
   \centering
   \includegraphics[width=0.9\columnwidth]{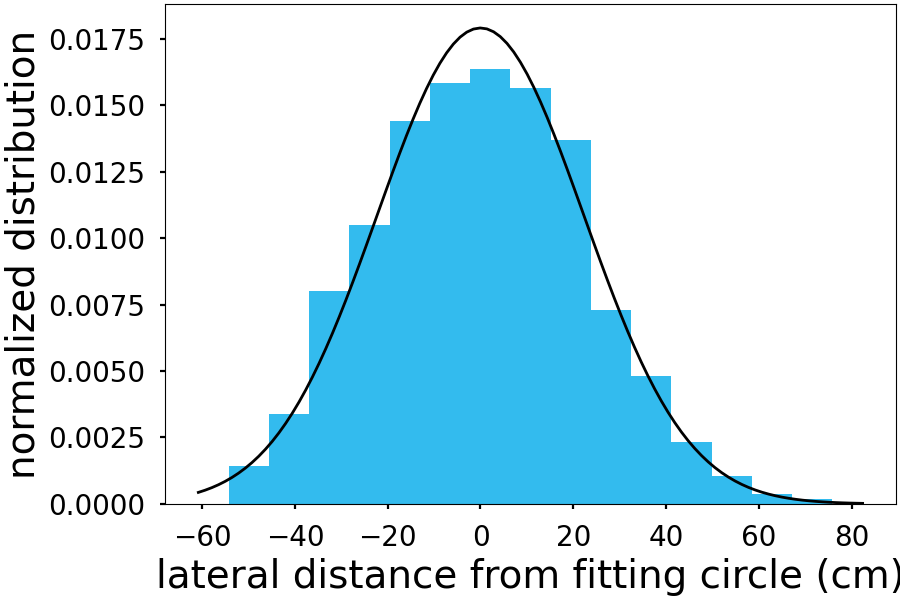}
   \caption{Distribution of the lateral distances from the fitting circle,
   averaged on all participants and all times, for the longest experiment.
   The distribution is fitted by a Gaussian of standard deviation $\sigma = 22$ cm.}
   \label{fig_latd}
\end{figure}

To quantify whether participants were able to form a circle,
we measured the lateral distance of each participant from the
participant from the matching circle at each point in time, and found 
that the distribution of these distances is typically Gaussian with a standard deviation of $\sigma = 0.19 \pm 2$ m (Fig.~\ref{fig_latd} 
shows the distribution for one of the realizations\redw{, more can be found in section~IX of the Suppl. Mat.}). 
This value can be compared with the average radius of the circle, which is $R = 3.8 \pm 0.2$ meters. So we find that $\sigma \simeq 5$ \% of $R$.

Assuming that the center of the circle follows a random walk,
we have calculated its diffusion coefficient from the relation valid in 2 dimensions
\begin{equation}
< r_c^2>(t) = 4Dt
\end{equation}
where $r_c$ is the distance of the center from its initial position.
The resulting diffusion coefficient is $D \simeq 30$ $cm^2/s$, which means that after
a minute, the typical displacement of the center will be $\sqrt{< r_c^2>} \simeq 80$ cm,
or, in other words, $20 \%$ of the radius.

\begin{figure}[ht] 
   \centering
   \includegraphics[width=0.9\columnwidth]{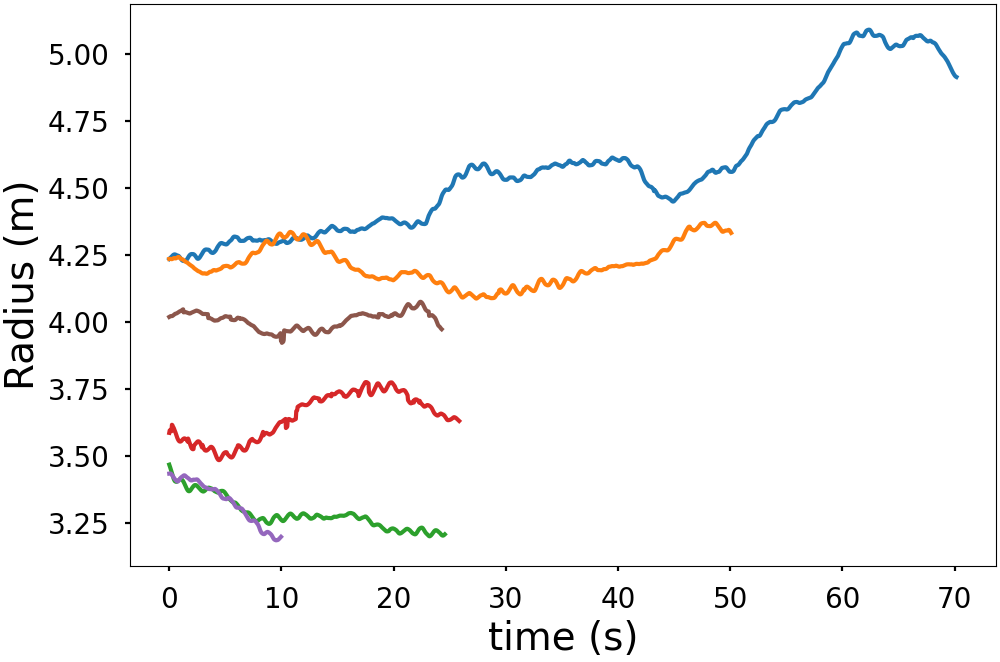}
   \caption{Radius of the fitting circle as a function of time, for each realization
   of the circle experiment.
   }
   \label{fig_radiusc}
\end{figure}

We have seen that participants are quite successful at forming circles. However, the radius of the circle can be quite different depending on the experiment, while it is quite stable in a given realization.
So it seems that the state reached is not always exactly the same, but once chosen, participants stick to it as if it were the result of a collective decision.

It is therefore interesting to look more closely at how the circle is formed.

\begin{figure}[bht]  \includegraphics[width=0.44\columnwidth]{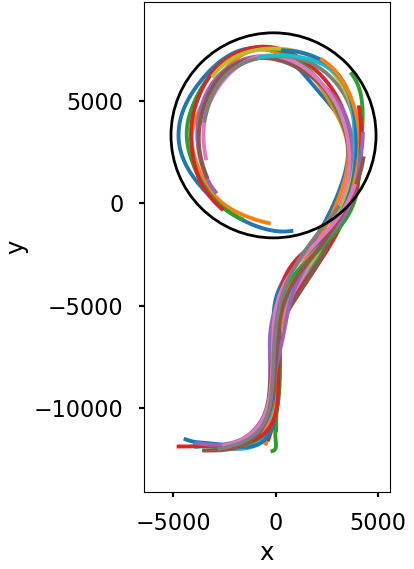}
   \hskip 0.05cm
   \includegraphics[width=0.54\columnwidth]{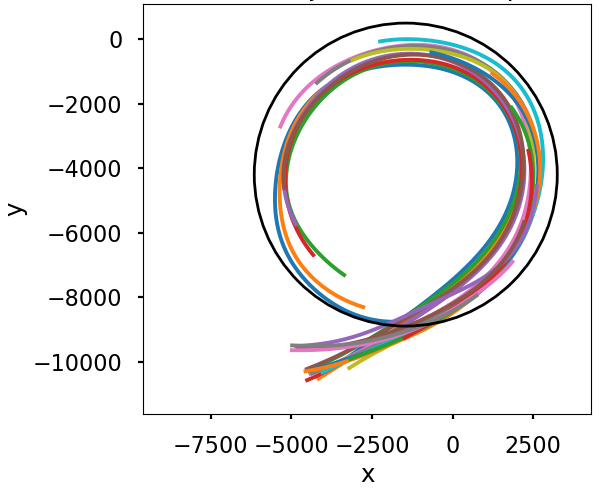} 
   \vskip 0.4cm
   \includegraphics[width=0.55\columnwidth]{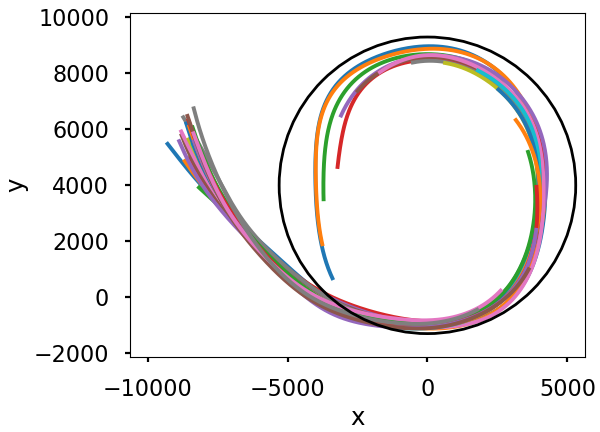}

   \caption{
   Trajectories of all participants during circle formation, for three different realizations. A circular area -black circle- called ``formation area'' is drawn by hand
   around the location of the future circle. 
   }
   \label{fig_circle_formation}
\end{figure}

Fig.~\ref{fig_circle_formation} shows the trajectories of all the participants during the formation of the circle. There is a remarkable overlap between them, which means on the one hand that the participants first form a line which they close by following each other, rather than coming from all directions, and on the other hand that the choice of the circle shape is anticipated long before the circle is completely closed.

\begin{figure}[ht] 
   \centering
\includegraphics[width=0.9\columnwidth]{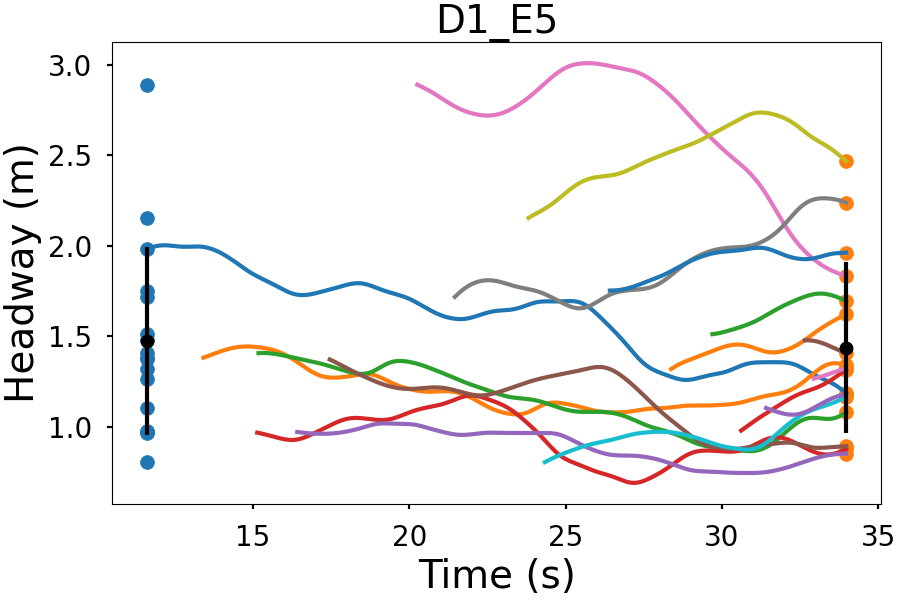}
 
\includegraphics[width=0.8\columnwidth]{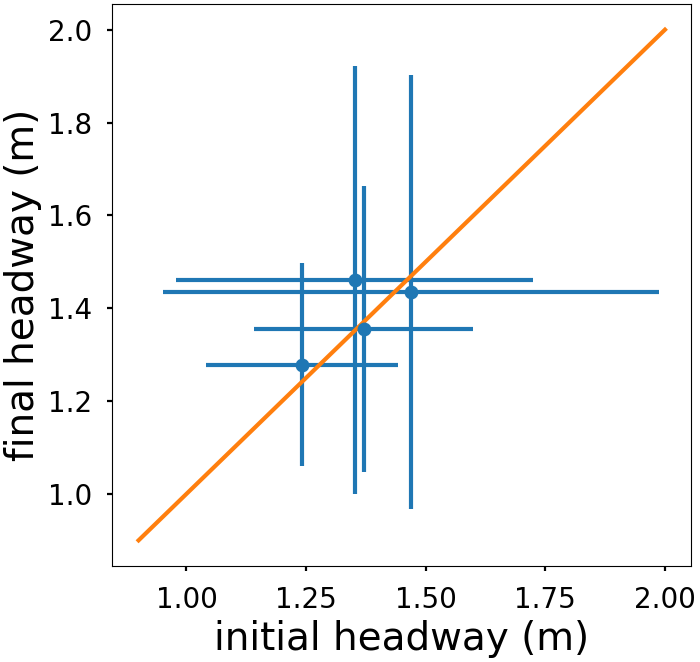}
   \caption{(Top) Distance headway of individual participants for one given realization, from the moment the first participant enter the formation area, up to the moment the last participant enters it. 
   The corresponding formation area is shown in Fig.~\ref{fig_circle_formation}-topleft. Dots indicate the headway values when entering the area (blue dots) or when the circle is finally formed (orange dots). (Bottom) Initial vs final headways, averaged over all participants, where {\em initial} refers to the moment at which participants enter the formation area of Fig.~\ref{fig_circle_formation}, and {\em final} to the moment where the circle is formed. Each point corresponds to a different realization of the experiment, and the orange line indicates the first diagonal as a guide to the eyes. Error bars give the standard deviation of individual values.
   }
   \label{fig_circle_formation_hh}
\end{figure}

To quantify this, we graphically define by hand a circular area within which the circle is formed. In the remaining of the paper, we call this area  ``formation area'' 
(limited by the black circle in Fig.~\ref{fig_circle_formation}).
For each participant, the headway is measured, first, when he or she enters the formation area, and second, when the circle is finally formed (blue and orange dots in Fig.~\ref{fig_circle_formation_hh}-top; see Section~\redw{VIII} of Suppl. Mat. for other realizations).
Though there are fluctuations, still the interdistances are relatively stable during the circle formation process.
This is confirmed by averaging the initial and final headways over all participants. Indeed, plotting final versus initial headways shows a good correlation (Fig.~\ref{fig_circle_formation_hh}-bottom). The only point slightly off the diagonal corresponds to the unique realization 
where one of the participants still had a significant headway relaxation after the circle was closed.

In order for participants to make this early choice of headways, a visual negotiation probably takes place, which is supported by the observation that many participants turn their head towards the participants on the opposite side while walking.

\section{Discussion and conclusion}
\label{sect_conclusion}

This paper presents a series of experiments on one-dimensional pedestrian flows.
In contrast to our previous experiments~\cite{jelic2012a}, here the flow
is less constrained, either because the participants can choose their path
( circle experiments), or because the absence
of a periodic boundary condition allows a free choice of headway.
In both cases, the participants can determine, individually or collectively, their density.
In addition, in the circle experiment, they can collectively choose their velocity.
This degree of freedom is only partially available in the line experiment.

We have shown that participants are quite at ease with the task of
forming a circle.
Remarkably, this leads to a very stable choice of speed modulus, possibly because the necessary collective choice plays the same role as a kind of averaging.

When the leader imposes a low upper limit on the speed (\textit{line-slow} experiments),
we found that the density is spontaneously increased to stay at the same
to the same operating point as when the density was imposed and the velocity was left free as in~\cite{jelic2012a}.
We could therefore say that {\em "imposed velocity"} or {\em "imposed density"} commute
in flows close to the jamming.
It would be interesting to check whether this is still the case for even lower leader velocities.

\redw{
By contrast, the line-free experimental results differ from what was observed in the constrained case of~\cite{jelic2012a}. Indeed, while here pedestrians would be free to choose large and comfortable headways, they keep instead relatively short distances combined with fairly high speeds.
}

\redw{
This lead us to one of our main results (see Fig.~\ref{fig_vh}, namely that in all the experiments in this paper, participants are keeping up with the fast adaptation branch that had only been observed at high densities in previous experiments.
}

\redw{
One possible explanation is that participants feel more social pressure when they are responsible for their decisions.
This could explain why they stick to behaviors that correspond to shorter adaptation times,
even at lower densities.
In contrast, in~\cite{jelic2012a} the movement is more constrained (fixed density, fixed path),
and participants are more passive.
They may then favor more comfort by choosing a behavior that allows a longer reaction time.
}

\redw{
As a conclusion, the type of constraints and the number of degrees of freedom may lead to different dynamical state. Our results suggest that one possible classification, in the spirit of ~\cite{cordes_s_n2024}, may rely on adaptation time scales, but further more systematic experiments would be needed.
}

\redw{In the general debate whether
the characteristics of the individual movement are mainly the result of the influence of the other individuals in the group (does the individual conform to the others), or whether each individual favours his own comfort of locomotion (with headways and speeds specific to his locomotor capacities), our results underline the importance of social pressure, and seem to indicate that, beyond the local information coming from the predecessor, pedestrians are aware of a more global dynamical state.}

It would be interesting to design new experiments to further disentangle the different effects of social pressure from behind,
comfort, goal attainment, locomotor constraints, step synchronisation, etc.
\redw{One could, for example, have individuals or groups carry out tracking tasks and compare their characteristics. This would be an original experiment exploring how individual movement characteristics are influenced by the group, whereas most existing experiments explore the emergence of macroscopic characteristics resulting from the combination of individual behaviors.}

\begin{acknowledgments}

We acknowledge some preliminary analysis of the data by Aymeric Duigou-Majumdar with financial support from Labex PALM (ANR-10-LABX-0039-PALM)
in the frame of "Investissements d'Avenir".

C. A.-R. acknowledges travel support from the project GAMEPED funded
by CNRS in the frame of the "D\'efi S2C3".

Experiments were designed and organized by the PEDINTERACT
project and realized at the laboratory M2S \redw{(Movement, Sport and health Sciences)} from Rennes 2.
We are in particular
grateful to Armel Cr\'etual and Anthony Sorel
for their help during the experiments.

PEDINTERACT experiments were funded by the Inria SIMS associated team as well as the ANR PERCOLATION ANR-13-JS02-0008 project.

\end{acknowledgments}

\bibliographystyle{apsrev4-2}

\begin{thebibliography}{37}%
\makeatletter
\providecommand \@ifxundefined [1]{%
 \@ifx{#1\undefined}
}%
\providecommand \@ifnum [1]{%
 \ifnum #1\expandafter \@firstoftwo
 \else \expandafter \@secondoftwo
 \fi
}%
\providecommand \@ifx [1]{%
 \ifx #1\expandafter \@firstoftwo
 \else \expandafter \@secondoftwo
 \fi
}%
\providecommand \natexlab [1]{#1}%
\providecommand \enquote  [1]{``#1''}%
\providecommand \bibnamefont  [1]{#1}%
\providecommand \bibfnamefont [1]{#1}%
\providecommand \citenamefont [1]{#1}%
\providecommand \href@noop [0]{\@secondoftwo}%
\providecommand \href [0]{\begingroup \@sanitize@url \@href}%
\providecommand \@href[1]{\@@startlink{#1}\@@href}%
\providecommand \@@href[1]{\endgroup#1\@@endlink}%
\providecommand \@sanitize@url [0]{\catcode `\\12\catcode `\$12\catcode `\&12\catcode `\#12\catcode `\^12\catcode `\_12\catcode `\%12\relax}%
\providecommand \@@startlink[1]{}%
\providecommand \@@endlink[0]{}%
\providecommand \url  [0]{\begingroup\@sanitize@url \@url }%
\providecommand \@url [1]{\endgroup\@href {#1}{\urlprefix }}%
\providecommand \urlprefix  [0]{URL }%
\providecommand \Eprint [0]{\href }%
\providecommand \doibase [0]{https://doi.org/}%
\providecommand \selectlanguage [0]{\@gobble}%
\providecommand \bibinfo  [0]{\@secondoftwo}%
\providecommand \bibfield  [0]{\@secondoftwo}%
\providecommand \translation [1]{[#1]}%
\providecommand \BibitemOpen [0]{}%
\providecommand \bibitemStop [0]{}%
\providecommand \bibitemNoStop [0]{.\EOS\space}%
\providecommand \EOS [0]{\spacefactor3000\relax}%
\providecommand \BibitemShut  [1]{\csname bibitem#1\endcsname}%
\let\auto@bib@innerbib\@empty
\bibitem [{\citenamefont {Haghani}(2020)}]{haghani2020empirical}%
  \BibitemOpen
  \bibfield  {author} {\bibinfo {author} {\bibfnamefont {M.}~\bibnamefont {Haghani}},\ }\href@noop {} {\bibfield  {journal} {\bibinfo  {journal} {Safety science}\ }\textbf {\bibinfo {volume} {129}},\ \bibinfo {pages} {104743} (\bibinfo {year} {2020})}\BibitemShut {NoStop}%
\bibitem [{\citenamefont {Corbetta}\ and\ \citenamefont {Toschi}(2023)}]{corbetta_t2023}%
  \BibitemOpen
  \bibfield  {author} {\bibinfo {author} {\bibfnamefont {A.}~\bibnamefont {Corbetta}}\ and\ \bibinfo {author} {\bibfnamefont {F.}~\bibnamefont {Toschi}},\ }\href {https://doi.org/10.1146/annurev-conmatphys-031620-100450} {\bibfield  {journal} {\bibinfo  {journal} {Annual Review of Condensed Matter Physics}\ }\textbf {\bibinfo {volume} {14}},\ \bibinfo {pages} {311} (\bibinfo {year} {2023})}\BibitemShut {NoStop}%
\bibitem [{\citenamefont {Ondrej}\ \emph {et~al.}(2010)\citenamefont {Ondrej}, \citenamefont {Pettr\'e}, \citenamefont {Olivier},\ and\ \citenamefont {Donikian}}]{ondrej2010b}%
  \BibitemOpen
  \bibfield  {author} {\bibinfo {author} {\bibfnamefont {J.}~\bibnamefont {Ondrej}}, \bibinfo {author} {\bibfnamefont {J.}~\bibnamefont {Pettr\'e}}, \bibinfo {author} {\bibfnamefont {A.-H.}\ \bibnamefont {Olivier}},\ and\ \bibinfo {author} {\bibfnamefont {S.}~\bibnamefont {Donikian}},\ }\href {https://doi.org/10.1145/1778765.1778860} {\bibfield  {journal} {\bibinfo  {journal} {ACM Trans. Graphics}\ }\textbf {\bibinfo {volume} {29}} (\bibinfo {year} {2010})}\BibitemShut {NoStop}%
\bibitem [{\citenamefont {Subaih}\ \emph {et~al.}(2024)\citenamefont {Subaih}, \citenamefont {Tordeux},\ and\ \citenamefont {Chraibi}}]{subaih_t_c2024}%
  \BibitemOpen
  \bibfield  {author} {\bibinfo {author} {\bibfnamefont {R.}~\bibnamefont {Subaih}}, \bibinfo {author} {\bibfnamefont {A.}~\bibnamefont {Tordeux}},\ and\ \bibinfo {author} {\bibfnamefont {M.}~\bibnamefont {Chraibi}},\ }\href {https://doi.org/10.17815/CD.2024.185} {\bibfield  {journal} {\bibinfo  {journal} {Collective Dynamics}\ }\textbf {\bibinfo {volume} {9}},\ \bibinfo {pages} {1} (\bibinfo {year} {2024})}\BibitemShut {NoStop}%
\bibitem [{\citenamefont {Lemercier}\ \emph {et~al.}(2012)\citenamefont {Lemercier}, \citenamefont {Jelic}, \citenamefont {Kulpa}, \citenamefont {Hua}, \citenamefont {Fehrenbach}, \citenamefont {Degond}, \citenamefont {Appert-Rolland}, \citenamefont {Donikian},\ and\ \citenamefont {Pettr\'e}}]{lemercier2012a}%
  \BibitemOpen
  \bibfield  {author} {\bibinfo {author} {\bibfnamefont {S.}~\bibnamefont {Lemercier}}, \bibinfo {author} {\bibfnamefont {A.}~\bibnamefont {Jelic}}, \bibinfo {author} {\bibfnamefont {R.}~\bibnamefont {Kulpa}}, \bibinfo {author} {\bibfnamefont {J.}~\bibnamefont {Hua}}, \bibinfo {author} {\bibfnamefont {J.}~\bibnamefont {Fehrenbach}}, \bibinfo {author} {\bibfnamefont {P.}~\bibnamefont {Degond}}, \bibinfo {author} {\bibfnamefont {C.}~\bibnamefont {Appert-Rolland}}, \bibinfo {author} {\bibfnamefont {S.}~\bibnamefont {Donikian}},\ and\ \bibinfo {author} {\bibfnamefont {J.}~\bibnamefont {Pettr\'e}},\ }\href {https://doi.org/10.1111/j.1467-8659.2012.03028.x} {\bibfield  {journal} {\bibinfo  {journal} {Computer Graphics Forum}\ }\textbf {\bibinfo {volume} {31}},\ \bibinfo {pages} {489} (\bibinfo {year} {2012})}\BibitemShut {NoStop}%
\bibitem [{\citenamefont {Seyfried}\ \emph {et~al.}(2005)\citenamefont {Seyfried}, \citenamefont {Steffen}, \citenamefont {Klingsch},\ and\ \citenamefont {Boltes}}]{seyfried2005}%
  \BibitemOpen
  \bibfield  {author} {\bibinfo {author} {\bibfnamefont {A.}~\bibnamefont {Seyfried}}, \bibinfo {author} {\bibfnamefont {B.}~\bibnamefont {Steffen}}, \bibinfo {author} {\bibfnamefont {W.}~\bibnamefont {Klingsch}},\ and\ \bibinfo {author} {\bibfnamefont {M.}~\bibnamefont {Boltes}},\ }\href@noop {} {\bibfield  {journal} {\bibinfo  {journal} {J. Stat. Mech.}\ ,\ \bibinfo {pages} {P10002}} (\bibinfo {year} {2005})}\BibitemShut {NoStop}%
\bibitem [{\citenamefont {Seyfried}\ \emph {et~al.}(2010{\natexlab{a}})\citenamefont {Seyfried}, \citenamefont {Boltes}, \citenamefont {Kahler}, \citenamefont {Klingsch}, \citenamefont {Portz}, \citenamefont {Rupprecht}, \citenamefont {Schadschneider}, \citenamefont {Steffen},\ and\ \citenamefont {Winkens}}]{seyfried2010}%
  \BibitemOpen
  \bibfield  {author} {\bibinfo {author} {\bibfnamefont {A.}~\bibnamefont {Seyfried}}, \bibinfo {author} {\bibfnamefont {M.}~\bibnamefont {Boltes}}, \bibinfo {author} {\bibfnamefont {J.}~\bibnamefont {Kahler}}, \bibinfo {author} {\bibfnamefont {W.}~\bibnamefont {Klingsch}}, \bibinfo {author} {\bibfnamefont {A.}~\bibnamefont {Portz}}, \bibinfo {author} {\bibfnamefont {T.}~\bibnamefont {Rupprecht}}, \bibinfo {author} {\bibfnamefont {A.}~\bibnamefont {Schadschneider}}, \bibinfo {author} {\bibfnamefont {B.}~\bibnamefont {Steffen}},\ and\ \bibinfo {author} {\bibfnamefont {A.}~\bibnamefont {Winkens}},\ }in\ \href {https://doi.org/10.1007/978-3-642-04504-2_11} {\emph {\bibinfo {booktitle} {Pedestrian and evacuation dynamics 2008}}},\ \bibinfo {editor} {edited by\ \bibinfo {editor} {\bibfnamefont {W.}~\bibnamefont {Klingsch}}, \bibinfo {editor} {\bibfnamefont {A.}~\bibnamefont {Schadschneider}},\ and\ \bibinfo {editor} {\bibfnamefont {M.}~\bibnamefont {Schreckenberg}}}\ (\bibinfo  {publisher} {Springer-Verlag
  Berlin},\ \bibinfo {year} {2010})\ pp.\ \bibinfo {pages} {145--156}\BibitemShut {NoStop}%
\bibitem [{\citenamefont {Cordes}\ \emph {et~al.}(2023)\citenamefont {Cordes}, \citenamefont {Chraibi}, \citenamefont {Tordeux},\ and\ \citenamefont {Schadschneider}}]{cordes2023}%
  \BibitemOpen
  \bibfield  {author} {\bibinfo {author} {\bibfnamefont {J.}~\bibnamefont {Cordes}}, \bibinfo {author} {\bibfnamefont {M.}~\bibnamefont {Chraibi}}, \bibinfo {author} {\bibfnamefont {A.}~\bibnamefont {Tordeux}},\ and\ \bibinfo {author} {\bibfnamefont {A.}~\bibnamefont {Schadschneider}},\ }in\ \href {https://doi.org/10.1007/978-3-031-46359-4_6} {\emph {\bibinfo {booktitle} {Crowd Dynamics, Volume 4: Analytics and Human Factors in Crowd Modeling}}},\ \bibinfo {editor} {edited by\ \bibinfo {editor} {\bibfnamefont {N.}~\bibnamefont {Bellomo}}\ and\ \bibinfo {editor} {\bibfnamefont {L.}~\bibnamefont {Gibelli}}}\ (\bibinfo  {publisher} {Birkh{\"a}user, Cham},\ \bibinfo {year} {2023})\ pp.\ \bibinfo {pages} {143--178}\BibitemShut {NoStop}%
\bibitem [{\citenamefont {Jeli\'c}\ \emph {et~al.}(2012{\natexlab{a}})\citenamefont {Jeli\'c}, \citenamefont {Appert-Rolland}, \citenamefont {Lemercier},\ and\ \citenamefont {Pettr\'e}}]{jelic2012a}%
  \BibitemOpen
  \bibfield  {author} {\bibinfo {author} {\bibfnamefont {A.}~\bibnamefont {Jeli\'c}}, \bibinfo {author} {\bibfnamefont {C.}~\bibnamefont {Appert-Rolland}}, \bibinfo {author} {\bibfnamefont {S.}~\bibnamefont {Lemercier}},\ and\ \bibinfo {author} {\bibfnamefont {J.}~\bibnamefont {Pettr\'e}},\ }\href {https://doi.org/10.1103/PhysRevE.85.036111} {\bibfield  {journal} {\bibinfo  {journal} {Phys. Rev. E}\ }\textbf {\bibinfo {volume} {85}},\ \bibinfo {pages} {036111} (\bibinfo {year} {2012}{\natexlab{a}})}\BibitemShut {NoStop}%
\bibitem [{\citenamefont {Cao}\ \emph {et~al.}(2016)\citenamefont {Cao}, \citenamefont {Zhang}, \citenamefont {Salden}, \citenamefont {Ma}, \citenamefont {Shi},\ and\ \citenamefont {Zhang}}]{cao2016a}%
  \BibitemOpen
  \bibfield  {author} {\bibinfo {author} {\bibfnamefont {S.}~\bibnamefont {Cao}}, \bibinfo {author} {\bibfnamefont {J.}~\bibnamefont {Zhang}}, \bibinfo {author} {\bibfnamefont {D.}~\bibnamefont {Salden}}, \bibinfo {author} {\bibfnamefont {J.}~\bibnamefont {Ma}}, \bibinfo {author} {\bibfnamefont {C.}~\bibnamefont {Shi}},\ and\ \bibinfo {author} {\bibfnamefont {R.}~\bibnamefont {Zhang}},\ }\href {https://doi.org/10.1103/PhysRevE.94.012312} {\bibfield  {journal} {\bibinfo  {journal} {Phys. Rev. E}\ }\textbf {\bibinfo {volume} {94}},\ \bibinfo {pages} {012312} (\bibinfo {year} {2016})}\BibitemShut {NoStop}%
\bibitem [{\citenamefont {Ren}\ \emph {et~al.}(2019)\citenamefont {Ren}, \citenamefont {Zhang}, \citenamefont {Song},\ and\ \citenamefont {Cao}}]{ren2019fundamental}%
  \BibitemOpen
  \bibfield  {author} {\bibinfo {author} {\bibfnamefont {X.}~\bibnamefont {Ren}}, \bibinfo {author} {\bibfnamefont {J.}~\bibnamefont {Zhang}}, \bibinfo {author} {\bibfnamefont {W.}~\bibnamefont {Song}},\ and\ \bibinfo {author} {\bibfnamefont {S.}~\bibnamefont {Cao}},\ }\href@noop {} {\bibfield  {journal} {\bibinfo  {journal} {Journal of Statistical Mechanics: Theory and Experiment}\ }\textbf {\bibinfo {volume} {2019}},\ \bibinfo {pages} {023403} (\bibinfo {year} {2019})}\BibitemShut {NoStop}%
\bibitem [{\citenamefont {Fang}\ \emph {et~al.}(2019)\citenamefont {Fang}, \citenamefont {Jiang}, \citenamefont {Li}, \citenamefont {Qi},\ and\ \citenamefont {Chen}}]{fang2019experimental}%
  \BibitemOpen
  \bibfield  {author} {\bibinfo {author} {\bibfnamefont {Z.~M.}\ \bibnamefont {Fang}}, \bibinfo {author} {\bibfnamefont {L.~X.}\ \bibnamefont {Jiang}}, \bibinfo {author} {\bibfnamefont {X.~L.}\ \bibnamefont {Li}}, \bibinfo {author} {\bibfnamefont {W.}~\bibnamefont {Qi}},\ and\ \bibinfo {author} {\bibfnamefont {L.~Z.}\ \bibnamefont {Chen}},\ }\href@noop {} {\bibfield  {journal} {\bibinfo  {journal} {Safety science}\ }\textbf {\bibinfo {volume} {113}},\ \bibinfo {pages} {264} (\bibinfo {year} {2019})}\BibitemShut {NoStop}%
\bibitem [{\citenamefont {Cao}\ \emph {et~al.}(2019)\citenamefont {Cao}, \citenamefont {Wang}, \citenamefont {Yao},\ and\ \citenamefont {Song}}]{cao2019dynamic}%
  \BibitemOpen
  \bibfield  {author} {\bibinfo {author} {\bibfnamefont {S.}~\bibnamefont {Cao}}, \bibinfo {author} {\bibfnamefont {P.}~\bibnamefont {Wang}}, \bibinfo {author} {\bibfnamefont {M.}~\bibnamefont {Yao}},\ and\ \bibinfo {author} {\bibfnamefont {W.}~\bibnamefont {Song}},\ }\href@noop {} {\bibfield  {journal} {\bibinfo  {journal} {Communications in Nonlinear Science and Numerical Simulation}\ }\textbf {\bibinfo {volume} {69}},\ \bibinfo {pages} {329} (\bibinfo {year} {2019})}\BibitemShut {NoStop}%
\bibitem [{\citenamefont {Chen}\ \emph {et~al.}(2017)\citenamefont {Chen}, \citenamefont {Lo},\ and\ \citenamefont {Ma}}]{chen2017pedestrian}%
  \BibitemOpen
  \bibfield  {author} {\bibinfo {author} {\bibfnamefont {J.}~\bibnamefont {Chen}}, \bibinfo {author} {\bibfnamefont {S.}~\bibnamefont {Lo}},\ and\ \bibinfo {author} {\bibfnamefont {J.}~\bibnamefont {Ma}},\ }\href@noop {} {\bibfield  {journal} {\bibinfo  {journal} {Journal of Statistical Mechanics: Theory and Experiment}\ }\textbf {\bibinfo {volume} {2017}},\ \bibinfo {pages} {083403} (\bibinfo {year} {2017})}\BibitemShut {NoStop}%
\bibitem [{\citenamefont {Sun}\ \emph {et~al.}(2018)\citenamefont {Sun}, \citenamefont {Lu}, \citenamefont {Lo}, \citenamefont {Ma},\ and\ \citenamefont {Xie}}]{sun2018moving}%
  \BibitemOpen
  \bibfield  {author} {\bibinfo {author} {\bibfnamefont {J.}~\bibnamefont {Sun}}, \bibinfo {author} {\bibfnamefont {S.}~\bibnamefont {Lu}}, \bibinfo {author} {\bibfnamefont {S.}~\bibnamefont {Lo}}, \bibinfo {author} {\bibfnamefont {J.}~\bibnamefont {Ma}},\ and\ \bibinfo {author} {\bibfnamefont {Q.}~\bibnamefont {Xie}},\ }\href@noop {} {\bibfield  {journal} {\bibinfo  {journal} {Physica A: Statistical Mechanics and its Applications}\ }\textbf {\bibinfo {volume} {490}},\ \bibinfo {pages} {476} (\bibinfo {year} {2018})}\BibitemShut {NoStop}%
\bibitem [{\citenamefont {Huang}\ \emph {et~al.}(2018)\citenamefont {Huang}, \citenamefont {Zhang}, \citenamefont {Lo}, \citenamefont {Lu},\ and\ \citenamefont {Li}}]{huang2018experimental}%
  \BibitemOpen
  \bibfield  {author} {\bibinfo {author} {\bibfnamefont {S.}~\bibnamefont {Huang}}, \bibinfo {author} {\bibfnamefont {T.}~\bibnamefont {Zhang}}, \bibinfo {author} {\bibfnamefont {S.}~\bibnamefont {Lo}}, \bibinfo {author} {\bibfnamefont {S.}~\bibnamefont {Lu}},\ and\ \bibinfo {author} {\bibfnamefont {C.}~\bibnamefont {Li}},\ }\href@noop {} {\bibfield  {journal} {\bibinfo  {journal} {Physica A: Statistical Mechanics and its Applications}\ }\textbf {\bibinfo {volume} {509}},\ \bibinfo {pages} {1023} (\bibinfo {year} {2018})}\BibitemShut {NoStop}%
\bibitem [{\citenamefont {Shi}\ \emph {et~al.}(2018)\citenamefont {Shi}, \citenamefont {Ye}, \citenamefont {Shiwakoti},\ and\ \citenamefont {Grembek}}]{shi2018state}%
  \BibitemOpen
  \bibfield  {author} {\bibinfo {author} {\bibfnamefont {X.}~\bibnamefont {Shi}}, \bibinfo {author} {\bibfnamefont {Z.}~\bibnamefont {Ye}}, \bibinfo {author} {\bibfnamefont {N.}~\bibnamefont {Shiwakoti}},\ and\ \bibinfo {author} {\bibfnamefont {O.}~\bibnamefont {Grembek}},\ }\href@noop {} {\bibfield  {journal} {\bibinfo  {journal} {Journal of Advanced Transportation}\ }\textbf {\bibinfo {volume} {2018}} (\bibinfo {year} {2018})}\BibitemShut {NoStop}%
\bibitem [{\citenamefont {Zeng}\ \emph {et~al.}(2019)\citenamefont {Zeng}, \citenamefont {Schadschneider}, \citenamefont {Zhang}, \citenamefont {Wei}, \citenamefont {Song},\ and\ \citenamefont {Ba}}]{zeng2019experimental}%
  \BibitemOpen
  \bibfield  {author} {\bibinfo {author} {\bibfnamefont {G.}~\bibnamefont {Zeng}}, \bibinfo {author} {\bibfnamefont {A.}~\bibnamefont {Schadschneider}}, \bibinfo {author} {\bibfnamefont {J.}~\bibnamefont {Zhang}}, \bibinfo {author} {\bibfnamefont {S.}~\bibnamefont {Wei}}, \bibinfo {author} {\bibfnamefont {W.}~\bibnamefont {Song}},\ and\ \bibinfo {author} {\bibfnamefont {R.}~\bibnamefont {Ba}},\ }\href@noop {} {\bibfield  {journal} {\bibinfo  {journal} {Physics Letters A}\ }\textbf {\bibinfo {volume} {383}},\ \bibinfo {pages} {1011} (\bibinfo {year} {2019})}\BibitemShut {NoStop}%
\bibitem [{\citenamefont {Yanagisawa}\ \emph {et~al.}(2012)\citenamefont {Yanagisawa}, \citenamefont {Tomoeda},\ and\ \citenamefont {Nishinari}}]{yanagisawa_t_n2012}%
  \BibitemOpen
  \bibfield  {author} {\bibinfo {author} {\bibfnamefont {D.}~\bibnamefont {Yanagisawa}}, \bibinfo {author} {\bibfnamefont {A.}~\bibnamefont {Tomoeda}},\ and\ \bibinfo {author} {\bibfnamefont {K.}~\bibnamefont {Nishinari}},\ }\href@noop {} {\bibfield  {journal} {\bibinfo  {journal} {Phys. Rev. E}\ }\textbf {\bibinfo {volume} {85}},\ \bibinfo {pages} {016111} (\bibinfo {year} {2012})}\BibitemShut {NoStop}%
\bibitem [{\citenamefont {Hu}\ \emph {et~al.}(2021)\citenamefont {Hu}, \citenamefont {Zhang}, \citenamefont {Song},\ and\ \citenamefont {Bode}}]{hu2021social}%
  \BibitemOpen
  \bibfield  {author} {\bibinfo {author} {\bibfnamefont {Y.}~\bibnamefont {Hu}}, \bibinfo {author} {\bibfnamefont {J.}~\bibnamefont {Zhang}}, \bibinfo {author} {\bibfnamefont {W.}~\bibnamefont {Song}},\ and\ \bibinfo {author} {\bibfnamefont {N.~W.}\ \bibnamefont {Bode}},\ }\href@noop {} {\bibfield  {journal} {\bibinfo  {journal} {Safety science}\ }\textbf {\bibinfo {volume} {139}},\ \bibinfo {pages} {105259} (\bibinfo {year} {2021})}\BibitemShut {NoStop}%
\bibitem [{\citenamefont {Chattaraj}\ \emph {et~al.}(2009)\citenamefont {Chattaraj}, \citenamefont {Seyfried},\ and\ \citenamefont {Chakroborty}}]{chattaraj_s_c2009a}%
  \BibitemOpen
  \bibfield  {author} {\bibinfo {author} {\bibfnamefont {U.}~\bibnamefont {Chattaraj}}, \bibinfo {author} {\bibfnamefont {A.}~\bibnamefont {Seyfried}},\ and\ \bibinfo {author} {\bibfnamefont {P.}~\bibnamefont {Chakroborty}},\ }\href@noop {} {\bibfield  {journal} {\bibinfo  {journal} {Advances in Complex Systems}\ }\textbf {\bibinfo {volume} {12}},\ \bibinfo {pages} {393} (\bibinfo {year} {2009})}\BibitemShut {NoStop}%
\bibitem [{\citenamefont {Gulhare}\ \emph {et~al.}(2018)\citenamefont {Gulhare}, \citenamefont {Verma},\ and\ \citenamefont {Chakroborty}}]{gulhare2018comparison}%
  \BibitemOpen
  \bibfield  {author} {\bibinfo {author} {\bibfnamefont {S.}~\bibnamefont {Gulhare}}, \bibinfo {author} {\bibfnamefont {A.}~\bibnamefont {Verma}},\ and\ \bibinfo {author} {\bibfnamefont {P.}~\bibnamefont {Chakroborty}},\ }\href@noop {} {\bibfield  {journal} {\bibinfo  {journal} {Collective Dynamics}\ }\textbf {\bibinfo {volume} {3}},\ \bibinfo {pages} {1} (\bibinfo {year} {2018})}\BibitemShut {NoStop}%
\bibitem [{\citenamefont {Zhang}\ and\ \citenamefont {Huang}(2024)}]{zhang_h2024}%
  \BibitemOpen
  \bibfield  {author} {\bibinfo {author} {\bibfnamefont {Y.}~\bibnamefont {Zhang}}\ and\ \bibinfo {author} {\bibfnamefont {X.}~\bibnamefont {Huang}},\ }\href {https://doi.org/10.1007/s10694-022-01357-5} {\bibfield  {journal} {\bibinfo  {journal} {Fire Technology}\ }\textbf {\bibinfo {volume} {60}},\ \bibinfo {pages} {859} (\bibinfo {year} {2024})}\BibitemShut {NoStop}%
\bibitem [{\citenamefont {Jeli\'c}\ \emph {et~al.}(2012{\natexlab{b}})\citenamefont {Jeli\'c}, \citenamefont {Appert-Rolland}, \citenamefont {Lemercier},\ and\ \citenamefont {Pettr\'e}}]{jelic2012b}%
  \BibitemOpen
  \bibfield  {author} {\bibinfo {author} {\bibfnamefont {A.}~\bibnamefont {Jeli\'c}}, \bibinfo {author} {\bibfnamefont {C.}~\bibnamefont {Appert-Rolland}}, \bibinfo {author} {\bibfnamefont {S.}~\bibnamefont {Lemercier}},\ and\ \bibinfo {author} {\bibfnamefont {J.}~\bibnamefont {Pettr\'e}},\ }\href@noop {} {\bibfield  {journal} {\bibinfo  {journal} {Phys. Rev. E}\ }\textbf {\bibinfo {volume} {86}},\ \bibinfo {pages} {046111} (\bibinfo {year} {2012}{\natexlab{b}})}\BibitemShut {NoStop}%
\bibitem [{\citenamefont {Zeng}\ \emph {et~al.}(2018)\citenamefont {Zeng}, \citenamefont {Cao}, \citenamefont {Liu},\ and\ \citenamefont {Song}}]{zeng2018experimental}%
  \BibitemOpen
  \bibfield  {author} {\bibinfo {author} {\bibfnamefont {G.}~\bibnamefont {Zeng}}, \bibinfo {author} {\bibfnamefont {S.}~\bibnamefont {Cao}}, \bibinfo {author} {\bibfnamefont {C.}~\bibnamefont {Liu}},\ and\ \bibinfo {author} {\bibfnamefont {W.}~\bibnamefont {Song}},\ }\href@noop {} {\bibfield  {journal} {\bibinfo  {journal} {Physica A: Statistical Mechanics and its Applications}\ }\textbf {\bibinfo {volume} {500}},\ \bibinfo {pages} {237} (\bibinfo {year} {2018})}\BibitemShut {NoStop}%
\bibitem [{\citenamefont {Cao}\ \emph {et~al.}(2018)\citenamefont {Cao}, \citenamefont {Zhang}, \citenamefont {Song}, \citenamefont {Zhang} \emph {et~al.}}]{cao2018stepping}%
  \BibitemOpen
  \bibfield  {author} {\bibinfo {author} {\bibfnamefont {S.}~\bibnamefont {Cao}}, \bibinfo {author} {\bibfnamefont {J.}~\bibnamefont {Zhang}}, \bibinfo {author} {\bibfnamefont {W.}~\bibnamefont {Song}}, \bibinfo {author} {\bibfnamefont {R.}~\bibnamefont {Zhang}}, \emph {et~al.},\ }\href@noop {} {\bibfield  {journal} {\bibinfo  {journal} {Journal of Statistical Mechanics: Theory and Experiment}\ }\textbf {\bibinfo {volume} {2018}},\ \bibinfo {pages} {033402} (\bibinfo {year} {2018})}\BibitemShut {NoStop}%
\bibitem [{\citenamefont {Ma}\ \emph {et~al.}(2018)\citenamefont {Ma}, \citenamefont {Sun}, \citenamefont {Lee},\ and\ \citenamefont {Yuen}}]{ma2018pedestrian}%
  \BibitemOpen
  \bibfield  {author} {\bibinfo {author} {\bibfnamefont {Y.}~\bibnamefont {Ma}}, \bibinfo {author} {\bibfnamefont {Y.~Y.}\ \bibnamefont {Sun}}, \bibinfo {author} {\bibfnamefont {E.~W.~M.}\ \bibnamefont {Lee}},\ and\ \bibinfo {author} {\bibfnamefont {R.~K.~K.}\ \bibnamefont {Yuen}},\ }\href@noop {} {\bibfield  {journal} {\bibinfo  {journal} {Physical Review E}\ }\textbf {\bibinfo {volume} {98}},\ \bibinfo {pages} {062311} (\bibinfo {year} {2018})}\BibitemShut {NoStop}%
\bibitem [{\citenamefont {Ren}\ \emph {et~al.}(2021)\citenamefont {Ren}, \citenamefont {Zhang},\ and\ \citenamefont {Song}}]{ren2021flows}%
  \BibitemOpen
  \bibfield  {author} {\bibinfo {author} {\bibfnamefont {X.}~\bibnamefont {Ren}}, \bibinfo {author} {\bibfnamefont {J.}~\bibnamefont {Zhang}},\ and\ \bibinfo {author} {\bibfnamefont {W.}~\bibnamefont {Song}},\ }\href@noop {} {\bibfield  {journal} {\bibinfo  {journal} {Safety Science}\ }\textbf {\bibinfo {volume} {133}},\ \bibinfo {pages} {105040} (\bibinfo {year} {2021})}\BibitemShut {NoStop}%
\bibitem [{\citenamefont {Yanagisawa}\ \emph {et~al.}(2013)\citenamefont {Yanagisawa}, \citenamefont {Suma}, \citenamefont {Tomoeda}, \citenamefont {Miura}, \citenamefont {Ohtsuka},\ and\ \citenamefont {Nishinari}}]{yanagisawa2013walking}%
  \BibitemOpen
  \bibfield  {author} {\bibinfo {author} {\bibfnamefont {D.}~\bibnamefont {Yanagisawa}}, \bibinfo {author} {\bibfnamefont {Y.}~\bibnamefont {Suma}}, \bibinfo {author} {\bibfnamefont {A.}~\bibnamefont {Tomoeda}}, \bibinfo {author} {\bibfnamefont {A.}~\bibnamefont {Miura}}, \bibinfo {author} {\bibfnamefont {K.}~\bibnamefont {Ohtsuka}},\ and\ \bibinfo {author} {\bibfnamefont {K.}~\bibnamefont {Nishinari}},\ }\href@noop {} {\bibfield  {journal} {\bibinfo  {journal} {Transportation research part C: emerging technologies}\ }\textbf {\bibinfo {volume} {37}},\ \bibinfo {pages} {238} (\bibinfo {year} {2013})}\BibitemShut {NoStop}%
\bibitem [{\citenamefont {Appert-Rolland}\ \emph {et~al.}(2020)\citenamefont {Appert-Rolland}, \citenamefont {Pettr\'e}, \citenamefont {Olivier}, \citenamefont {Warren}, \citenamefont {Duigou-Majumdar}, \citenamefont {Pinsard},\ and\ \citenamefont {Nicolas}}]{appert-rolland2020b}%
  \BibitemOpen
  \bibfield  {author} {\bibinfo {author} {\bibfnamefont {C.}~\bibnamefont {Appert-Rolland}}, \bibinfo {author} {\bibfnamefont {J.}~\bibnamefont {Pettr\'e}}, \bibinfo {author} {\bibfnamefont {A.-H.}\ \bibnamefont {Olivier}}, \bibinfo {author} {\bibfnamefont {W.}~\bibnamefont {Warren}}, \bibinfo {author} {\bibfnamefont {A.}~\bibnamefont {Duigou-Majumdar}}, \bibinfo {author} {\bibfnamefont {E.}~\bibnamefont {Pinsard}},\ and\ \bibinfo {author} {\bibfnamefont {A.}~\bibnamefont {Nicolas}},\ }\href {https://doi.org/10.17815/CD.2020.109} {\bibfield  {journal} {\bibinfo  {journal} {Collective Dynamics}\ }\textbf {\bibinfo {volume} {5}},\ \bibinfo {pages} {1} (\bibinfo {year} {2020})}\BibitemShut {NoStop}%
\bibitem [{\citenamefont {Lemercier}\ \emph {et~al.}(2011)\citenamefont {Lemercier}, \citenamefont {Moreau}, \citenamefont {Moussa{\"i}d}, \citenamefont {Theraulaz}, \citenamefont {Donikian},\ and\ \citenamefont {Pettr\'e}}]{lemercier2011b}%
  \BibitemOpen
  \bibfield  {author} {\bibinfo {author} {\bibfnamefont {S.}~\bibnamefont {Lemercier}}, \bibinfo {author} {\bibfnamefont {M.}~\bibnamefont {Moreau}}, \bibinfo {author} {\bibfnamefont {M.}~\bibnamefont {Moussa{\"i}d}}, \bibinfo {author} {\bibfnamefont {G.}~\bibnamefont {Theraulaz}}, \bibinfo {author} {\bibfnamefont {S.}~\bibnamefont {Donikian}},\ and\ \bibinfo {author} {\bibfnamefont {J.}~\bibnamefont {Pettr\'e}},\ }\href {https://doi.org/10.1007/978-3-642-25090-3_31} {\bibfield  {journal} {\bibinfo  {journal} {Lecture Notes in Computer Science}\ }\textbf {\bibinfo {volume} {7060}},\ \bibinfo {pages} {365} (\bibinfo {year} {2011})}\BibitemShut {NoStop}%
\bibitem [{Note1()}]{Note1}%
  \BibitemOpen
  \bibinfo {note} {Some authors define the density from a headway which is the average of the distance with the predecessor and the follower~\cite {seyfried2010}. Here, assuming that it is the distance {\protect \em ahead} of the participant that predominantly determines his speed, we have chosen to consider only the headway ahead.}\BibitemShut {Stop}%
\bibitem [{\citenamefont {Seyfried}\ \emph {et~al.}(2010{\natexlab{b}})\citenamefont {Seyfried}, \citenamefont {Portz},\ and\ \citenamefont {Schadschneider}}]{seyfried_p_s2010}%
  \BibitemOpen
  \bibfield  {author} {\bibinfo {author} {\bibfnamefont {A.}~\bibnamefont {Seyfried}}, \bibinfo {author} {\bibfnamefont {A.}~\bibnamefont {Portz}},\ and\ \bibinfo {author} {\bibfnamefont {A.}~\bibnamefont {Schadschneider}},\ }in\ \href@noop {} {\emph {\bibinfo {booktitle} {Lecture Notes in Computer Science}}},\ Vol.\ \bibinfo {volume} {6350},\ \bibinfo {editor} {edited by\ \bibinfo {editor} {\bibfnamefont {S.}~\bibnamefont {Bandini}}, \bibinfo {editor} {\bibfnamefont {S.}~\bibnamefont {Manzoni}}, \bibinfo {editor} {\bibfnamefont {H.}~\bibnamefont {Umeo}},\ and\ \bibinfo {editor} {\bibfnamefont {G.}~\bibnamefont {Vizzari}}}\ (\bibinfo  {publisher} {Springer-Verlag Berlin},\ \bibinfo {year} {2010})\ pp.\ \bibinfo {pages} {496--505}\BibitemShut {NoStop}%
\bibitem [{\citenamefont {Fehrenbach}\ \emph {et~al.}(2015)\citenamefont {Fehrenbach}, \citenamefont {Narski}, \citenamefont {Hua}, \citenamefont {Lemercier}, \citenamefont {Jeli\'c}, \citenamefont {Appert-Rolland}, \citenamefont {Donikian}, \citenamefont {Pettr\'e},\ and\ \citenamefont {Degond}}]{fehrenbach2015}%
  \BibitemOpen
  \bibfield  {author} {\bibinfo {author} {\bibfnamefont {J.}~\bibnamefont {Fehrenbach}}, \bibinfo {author} {\bibfnamefont {J.}~\bibnamefont {Narski}}, \bibinfo {author} {\bibfnamefont {J.}~\bibnamefont {Hua}}, \bibinfo {author} {\bibfnamefont {S.}~\bibnamefont {Lemercier}}, \bibinfo {author} {\bibfnamefont {A.}~\bibnamefont {Jeli\'c}}, \bibinfo {author} {\bibfnamefont {C.}~\bibnamefont {Appert-Rolland}}, \bibinfo {author} {\bibfnamefont {S.}~\bibnamefont {Donikian}}, \bibinfo {author} {\bibfnamefont {J.}~\bibnamefont {Pettr\'e}},\ and\ \bibinfo {author} {\bibfnamefont {P.}~\bibnamefont {Degond}},\ }\href {https://doi.org/10.3934/nhm.2015.10.579} {\bibfield  {journal} {\bibinfo  {journal} {Networks and Heterogeneous Media}\ }\textbf {\bibinfo {volume} {10}},\ \bibinfo {pages} {579} (\bibinfo {year} {2015})}\BibitemShut {NoStop}%
\bibitem [{\citenamefont {Nakayama}\ \emph {et~al.}(2016)\citenamefont {Nakayama}, \citenamefont {Kikuchi}, \citenamefont {Shibata}, \citenamefont {Sugiyama}, \citenamefont {i.~Tadaki},\ and\ \citenamefont {Yukawa}}]{nakayama2016}%
  \BibitemOpen
  \bibfield  {author} {\bibinfo {author} {\bibfnamefont {A.}~\bibnamefont {Nakayama}}, \bibinfo {author} {\bibfnamefont {M.}~\bibnamefont {Kikuchi}}, \bibinfo {author} {\bibfnamefont {A.}~\bibnamefont {Shibata}}, \bibinfo {author} {\bibfnamefont {Y.}~\bibnamefont {Sugiyama}}, \bibinfo {author} {\bibfnamefont {S.}~\bibnamefont {i.~Tadaki}},\ and\ \bibinfo {author} {\bibfnamefont {S.}~\bibnamefont {Yukawa}},\ }\href@noop {} {\bibfield  {journal} {\bibinfo  {journal} {New J. Phys.}\ }\textbf {\bibinfo {volume} {18}},\ \bibinfo {pages} {043040} (\bibinfo {year} {2016})}\BibitemShut {NoStop}%
\bibitem [{\citenamefont {Motsch}\ \emph {et~al.}(2018)\citenamefont {Motsch}, \citenamefont {Moussaid}, \citenamefont {Guillot}, \citenamefont {Moreau}, \citenamefont {Pettr\'e}, \citenamefont {Theraulaz}, \citenamefont {Appert-Rolland},\ and\ \citenamefont {Degond}}]{motsch2018}%
  \BibitemOpen
  \bibfield  {author} {\bibinfo {author} {\bibfnamefont {S.}~\bibnamefont {Motsch}}, \bibinfo {author} {\bibfnamefont {M.}~\bibnamefont {Moussaid}}, \bibinfo {author} {\bibfnamefont {E.}~\bibnamefont {Guillot}}, \bibinfo {author} {\bibfnamefont {M.}~\bibnamefont {Moreau}}, \bibinfo {author} {\bibfnamefont {J.}~\bibnamefont {Pettr\'e}}, \bibinfo {author} {\bibfnamefont {G.}~\bibnamefont {Theraulaz}}, \bibinfo {author} {\bibfnamefont {C.}~\bibnamefont {Appert-Rolland}},\ and\ \bibinfo {author} {\bibfnamefont {P.}~\bibnamefont {Degond}},\ }\href {https://doi.org/10.3934/mbe.2018059} {\bibfield  {journal} {\bibinfo  {journal} {Math. Biosci. Eng.}\ }\textbf {\bibinfo {volume} {15}},\ \bibinfo {pages} {1271} (\bibinfo {year} {2018})}\BibitemShut {NoStop}%
\bibitem [{\citenamefont {Cordes}\ \emph {et~al.}(2024)\citenamefont {Cordes}, \citenamefont {Schadschneider},\ and\ \citenamefont {Nicolas}}]{cordes_s_n2024}%
  \BibitemOpen
  \bibfield  {author} {\bibinfo {author} {\bibfnamefont {J.}~\bibnamefont {Cordes}}, \bibinfo {author} {\bibfnamefont {A.}~\bibnamefont {Schadschneider}},\ and\ \bibinfo {author} {\bibfnamefont {A.}~\bibnamefont {Nicolas}},\ }\href {https://doi.org/10.1093/pnasnexus/pgae120} {\bibfield  {journal} {\bibinfo  {journal} {PNAS Nexus}\ }\textbf {\bibinfo {volume} {3}},\ \bibinfo {pages} {pgae120} (\bibinfo {year} {2024})}\BibitemShut {NoStop}%
\end{thebibliography}

%

\end{document}